\documentclass[%
 reprint,
amsmath,amssymb,
aps,
prb,
]{revtex4-1}

\usepackage{graphicx}
\usepackage{color}
\usepackage{dcolumn}
\usepackage{multirow}

\usepackage{bm}
\usepackage{bbold}
\usepackage{dsfont}

\begin{document}

\preprint{APS/123-QED}

\title{Abelian Chern-Simons Theory for the Fractional Quantum Hall Effect in Graphene}

\author{Christian Fr\"a\ss dorf}
\affiliation{%
Dahlem Center for Complex Quantum Systems and, Institut f\"ur Theoretische Physik, \\Freie Universit\"at Berlin, Arnimallee 14, 14195 Berlin, Germany}%

\date{\today}%

\begin{abstract}
We develop a theory for the pseudorelativistic fractional quantum Hall effect in graphene, which is based on a multicomponent abelian Chern-Simons theory in the fermionic functional integral approach. Calculations are performed in the Keldysh formalism, directly giving access to real-time correlation functions at finite temperature. We obtain an exact effective action for the Chern-Simons gauge fields, which is expanded to second order in the gauge field fluctations around the mean-field solution. The one-loop fermionic polarization tensor as well as the electromagnetic response tensor in random phase approximation are derived, from which we obtain the Hall conductivities for various FQH states, lying symmetrically around charge neutrality.

\end{abstract}

\maketitle

\section{Introduction}
\label{sec:Introduction}
The integer quantum Hall effect (IQHE) is a remarkable experimental discovery of the early 1980s, since it proves quantum mechanics at work on macroscopic scales.~\cite{KlitzingEtal1980}
In a nonrelativistic two-dimensional electron gas (2DEG) at low temperatures and in high external magnetic fields, the Hall conductivity shows a plateau structure as a function of the magnetic field or chemical potential occuring at integer multiples of the ``conductance quantum'' $e^2/h$. Remarkably, the existence of these plateaus can already be understood in simple noninteracting models by the formation of discrete, equidistant energy levels, the Landau levels (LLs).~\cite{JainBook}

In sharp contrast to an ordinary 2DEG with its parabolic band structure, in the vicinity of the charge neutrality point the band structure of graphene mimics the energy-momentum dispersion of massless, relativistic Dirac particles.~\cite{Wallace1947, Semenoff1984, GusyninSharapovCarbotte2007, CastroNetoGuineaNovoselovGeim2009, Dassarma2011} When subjected to strong magnetic fields such a pseudorelativistic dispersion relation has profound consequences on the LLs, which, in turn, influences the measurable Hall conductivity.~\cite{CastroNetoGuineaNovoselovGeim2009, Goerbig2011} In theoretical studies one finds an anomalous quantization, where each of the four fermionic flavours in graphene contributes a half-integer, $n + 1/2$, to the total Hall conductivity~\cite{GorbarGusynin2002, GusyninSharapov2005, Goerbig2011}
\begin{equation}
\sigma_{0, xy} = \pm 4 \left(n + \frac{1}{2} \right) \frac{e^2}{h} \,, \quad n = 0, 1, 2, \ldots \,.
\label{eq:IntegerHallConductivityIntro}
\end{equation}
The additional fraction of $1/2$ can be traced back to the existence of a half-filled Landau level located directly at the charge neutrality point, which has only half the degeneracy of the other levels (the spectral anomaly), while the factor of four is a direct consequence of the four independent $SU(4)$ symmetric flavours of charge carriers in the low energy Dirac model. With the recent success of graphene's experimental isolation these theoretical predictions became experimentally accessible and could indeed be verified.~\cite{Zhang2005, Novoselov2007}

Shortly after the IQHE was discovered in nonrelativistic semiconducting devices, measurements on high quality samples revealed the occurence of additional plateaus at certain fractional fillings,~\cite{TsuiEtal1982, PanStormer2002} and more recently this effect has also been observed in graphene.~\cite{DuEtal2009, BolotinEtal2009, Dean2011} For this fractional quantum Hall effect (FQHE) electron-electron interactions are an essential ingredient in the theoretical treatment to gain further understanding of the underlying physics. The main difficulty here is that the noninteracting Landau levels, forming the basis of the analysis, are macroscopically degenerate. As a consequence conventional perturbative approaches inevitably fail, making the FQH system a prime example for strongly correlated matter, which has to be analyzed by truly nonperturbative methods. 

Based on the seminal work of Laughlin,~\cite{Laughlin1983} Jain introduced the idea that physical electrons/holes and magnetic flux quanta, or vortices, form bound states, so-called ``composite fermions''.~\cite{Jain1989} Due to the process of flux nucleation, the magnetic field is reduced, leading to a new set of effective Landau levels that are occupied by the composite fermions. The integer fillings of those effective LLs map to the fractional fillings observed in the experiments. Thus, the fractional QHE of ordinary fermions can be understood as an integer QHE of composite fermions.~\cite{JainBook, HeinonenBook} This intuitive, albeit rather unconventional picture led to a vast body of theoretical predictions, which could be verified experimentally to a large extent.~\cite{DuStormer1993, Willett1993, Kang1993, Manoharan1994, Goldmann1994, Du1994, Leadley1994, Smet1996, Smet1998, Smet1999, Willett1999, Melinte2000, Kukushin2007, Kamburov2012, Kamburov2013, Jain2014} Applying these ideas to the Dirac electrons in graphene leads to the notion of ``composite Dirac fermions''. Accordingly, one might expect that their pseudorelativistic spectrum, which leads to the anomalous quantization of the Hall conductivity in the noninteracting case, leaves its marks in the FQHE.

In the theoretical treatment of the FQHE there are several slightly different approaches to realize Jain's idea of flux-binding. Within the trial wavefunction approach vortices are attached in the form of Jastrow factors multiplying the many-body wavefunction of noninteracting fermions in an IQH state.~\cite{JainBook, HeinonenBook} To make use of this strategy in graphene one considers a completely empty or completely filled lowest LL - usually the one at the charge neutrality point - as the vacuum state and attaches flux quanta to the physical electrons/holes that partially fill/deplete this energy level. Thereby it is assumed that the effective LLs and their associated single-particle wavefunctions which make up the IQHS are \textit{not} of the Dirac type, but coincide with the nonrelativistic Schr\"odinger type ones.~\cite{ToekeJain2006, ToekeJain2007, Goerbig2011} This assumption is justified by the fact that the quenched Hamiltonian of graphene projected to the lowest LL is identical to the Hamiltonian encountered in systems with a nonrelativistic parabolic dispersion.~\cite{ApalkovChakraborty2006, ToekeJain2006, ToekeJain2007, Goerbig2011} Loosely speaking, graphene electrons confined to the lowest LL lose their identity as Dirac fermions upon projection, such that the only impact of graphene's unconventional band structure is the $SU(4)$ symmetry of the ansatz wavefunction, which derives from the $SU(4)$ symmetry of the individual fermionic flavours. This construction leads to the conventional Jain sequence and wavefunctions. Straightforward generalizations of this approach are given by Halperin wavefunctions,~\cite{Halperin1983, ScarolaJain2001, PapicEtal2009, PapicEtal2010, GoerbigRegnault2007, Goerbig2011} which potentially break the $SU(4)$ symmetry down to $SU(2)^{\otimes 2}$ or even $U(1)^{\otimes 4}$. 

Despite its indisputable successes the trial wavefunction approach has several drawbacks, two of which we want to comment on further. First, it crucially depends on projected Hamiltonians, which typically neglect LL mixing. While for nonrelativistic systems for the most part this is only a minor issue, since at large magnetic fields LL mixing is suppressed as $1/\sqrt{B}$,~\cite{SodemannMacDonald2013} in graphene it is a substantially more severe problem. Here, LL mixing is controlled by the fine structure constant $\alpha$, which is independent of the magnetic field and - more importantly - genuinely large ($\alpha \approx 2.2$ in suspended graphene), making LL mixing a nonperturbative problem already on the level of the Hamiltonian.~\cite{Peterson2013, Peterson2014} Hence, although the kinetic energy may be quenched within a partially filled LL, the electrons in graphene still feel their Dirac heritage. Yet, if LL mixing is taken into account at least perturbatively, Refs.~[\onlinecite{Peterson2013}] and [\onlinecite{Peterson2014}] reported - quite surprisingly - that it has practically no effect on the wavefunctions in the zeroth LL. Not entirely decoupled from the above, the second main problem is concerned with particle-hole symmetry, or rather its strong breaking inherent in the construction of trial wavefunctions. The origin of paticle-hole symmetry is different for nonrelativistic and relativistic systems. For the former it is only an emergent symmetry of the lowest LL projected Hamiltonian, but for the latter it is an exact symmetry of the unprojected Hamiltonian (and, hence, is a good symmetry even if LL mixing is taken into account). The construction of particle-hole conjugated wavefunctions is still possible, but the explicit symmetry breaking is not only unsatisfying but also comes with its own complications, see for example Refs.~[\onlinecite{Dyakonov2002}] and [\onlinecite{Son2015}] for a more elaborate discussion.

A complementary approach to the construction of explicit wavefunctions is the Chern-Simons field theory, which does not rely on a projection to the lowest LL. Here, magnetic flux tubes - which should be distinguished from the vortices of the wavefunction approach - are attached to the fermionic degrees of freedom either via a singular gauge transformation,~\cite{HalperinLeeRead1993, HeinonenBook} or equivalently via a minimal coupling of a Chern-Simons gauge field to the kinetic action in addition to a kinetic Chern-Simons term.~\cite{LopezFradkin1991, LopezFradkin1993, LopezFradkin1995, Zhang1995} (See also Ref.~[\onlinecite{ZhangHanssonKivelson1989}] for a similar treatment involving bosons.) In the process, ordinary fermions are transformed into composite fermions, whose nature - Schr\"odinger or Dirac - is determined by the structure of the kinetic action. Hence, as opposed to the picture drawn in Ref.~[\onlinecite{ToekeJain2006}], the Chern-Simons composite fermions in graphene are actual Dirac type particles. Accordingly one might expect that the spectral anomaly of the composite Dirac fermions (the half-integer quantization of the filling fractions) enters the analytical formulas for the total filling fraction/Hall conductivity of the electronic system. However, the graphene Chern-Simons theories proposed in Refs.~[\onlinecite{ModakEtal2011}] and [\onlinecite{Khveshchenko2007}] attach flux to the physical electrons/holes with respect to the bottom/top of the lowest LL, which eliminates the spectral anomaly and yields predictions for the total filling fraction that are in accordance with the wavefunction approach. Concerning LL mixing the Chern-Simons approaches reside on the other side of the spectrum, meaning there is a large amount of LL mixing,~\cite{FradkinBook} which is a result of the Chern-Simons transformation and the absence of projection. Regarding the nonperturbative nature of LL mixing in graphene this feature should not necessarily be considered a flaw, but the question remains, if the Chern-Simons induced LL mixing describes the physical reality accurately.

Although the non-Dirac nature of the composite fermions in graphene's lowest LL appears to be fully established by the results of Ref.~[\onlinecite{Peterson2014}], the conclusion that theoretical frameworks which employ Dirac type composite fermions, such as the aforementioned pseudorelativistic Chern-Simons theories of Refs.~[\onlinecite{ModakEtal2011}] and [\onlinecite{Khveshchenko2007}], lose their viability would be too hasty as Son's work, Ref.~[\onlinecite{Son2015}], impressively shows. Focussing on the conventional nonrelativistic FQH system, Son proposed a manifestly particle-hole symmetric, pseudorelativistic effective model, which declares Jain's composite fermion to be a Dirac particle by nature. Specifically, the $\nu = 1/2$ state is described by a charge neutral Dirac particle interacting with an emergent gauge field (\textit{not} of the Chern-Simons type), that forms a Fermi liquid, while Jain's principal sequence around half-filling can be explained as the IQHE of those Dirac quasiparticles, fully incorporating the particle-hole symmetry of the lowest Landau level.

In contrast to Son's effective model, in the present paper we employ a rather standard microscopic Chern-Simons theory, similar to Refs.~[\onlinecite{ModakEtal2011}] and [\onlinecite{Khveshchenko2007}]. The crucial difference to those works is the reference point at which we implement Chern-Simons flux-attachment, namely the particle-hole symmetric Dirac point at charge neutrality. This shift in the reference point should not be underestimated as a mere shift in the total filling fraction, since it allows for a flux-attachment scheme that is distinctively different from the aforementioned approaches. Instead of attaching flux to the physical electrons/holes, it is possible to attach flux to the charge carrier density, that is electron- or hole-like quasiparticles measured from the charge neutrality point. In particular, we obtain a mean-field equation which involves the charge carrier density, instead of the electron/hole density, and within the calculation of Gaussian fluctuations we naturally encounter pseudorelativistic propagators experiencing an effective magnetic field, which incorporate the spectral anomaly. Our central result is the electromagnetic polarization tensor in linear response to an external perturbation, which - among others - gives access to the Hall conductivity of the multicomponent fractional quantum Hall system
\begin{equation}
\sigma_{xy} = \sum_{\alpha} \sigma_{0, xy}^{\alpha} - \sum_{\alpha \beta} \sigma_{0, xy}^{\alpha} (\sigma_{0, xy} + \hat{\mathcal{K}}^{-1})_{\alpha \beta}^{-1} \sigma_{0, xy}^{\beta} \,.
\label{eq:FractionalHallConductivityIntro}
\end{equation}
Here, $\sigma_{0, xy}^{\alpha}$ is the Hall conductivity of a noninteracting, single flavour $\alpha$, which is half-integer quantized at low temperatures, due to the Dirac nature of the composite fermions, and $\hat{\mathcal{K}}$ is an integer-valued symmetric matrix accounting for the flux-attachment.~\cite{Note1}

We show that Eq.~\eqref{eq:FractionalHallConductivityIntro} leads to particle-hole symmetric Hall plateaus around charge neutrality, if positive flux-attachment to electron-like and negative flux-attachment to hole-like quasiparticles is considered. This observation enables us to construct manifestly particle-hole symmetric filling fractions as functions of the chemical potential. This result seems surprising, since the Chern-Simons term explicitly breaks particle-hole symmetry - which is why Son discarded such a term in his effective theory~\cite{Son2015} - irrespective of the reference point where the Chern-Simons flux is attached. Since this symmetry cannot be generated dynamically, particle-hole symmetric Hall plateaus are not expected to occur in such a symmetry broken theory. The puzzle is resolved as follows: The standard definition of the particle-hole transformation involves fermionic and bosonic fields only, but leaves the Chern-Simons coupling untouched. By allowing the coupling to depend on the sign of the carrier density we have altered the flux-attachment prescription in such a way to make it consistent with the standard particle-hole symmetry transformation. One may also interpret it the other way around: We use the standard flux-attachment but change the symmetry transformation to involve a sign flip of the Chern-Simons coupling. Thus, one may argue that the Chern-Simons term only breaks particle-hole symmetry in a weak sense, since it can be circumvented altogether by sufficiently modifying the symmetry transformation or the flux-attachment prescription, alleviating the seeming incompatibility of particle-hole symmetry and Chern-Simons theory. Furthermore, we show that the above formula reproduces the Hall conductivities proposed in Refs.~[\onlinecite{PeresEtal2006}] and [\onlinecite{JellalEtal2010}] as special cases, as well as several other filling fractions that have been obtained in the wavefunction approach.

In this paper we employ the real-time Keldysh formalism, which offers several technical advantages in comparison to the conventional real-time ground state formalism. This formulation will allow for a natural regularization of the otherwise ill-defined mean field equations, upon which the flux-attachment interpretation is based, and it additionally yields results that are valid at finite temperature, which come without further calculational costs. Our exposition is inspired by the original work of Refs.~[\onlinecite{LopezFradkin1991}] and [\onlinecite{LopezFradkin1993}], where the fermion Chern-Simons theory for the FQHE of nonrelativistic matter has been introduced. Since there are several subtle differences due to the Dirac nature of the quasiparticles and the Keldysh formulation, we will present the theory in a self-contained manner. The outline of the article is as follows: In section \ref{sec:AbelianChernSimonsTheory} we describe the field theory of interacting Dirac fermions coupled to statistical Chern-Simons fields with the abelian gauge group $U(1)^{\otimes 4}$. In the subsequent section we derive an exact effective action for the statistical gauge fields and discuss its Gaussian approximation around the mean field solution of the quantum Hall liquid. Section \ref{sec:ElectromagneticResponseTensorAndHallConductance} contains our main results. We address the topic of gauge fixing and calculate the full electromagnetic response tensor together with Hall conductivities for a selected set of states. We conclude in the final section. Further technical details of the computation are given in two appendices.

\section{Abelian Chern-Simons Theory}
\label{sec:AbelianChernSimonsTheory}
The starting point of our considerations is the second quantized low energy Hamiltonian for interacting Dirac electrons in monolayer graphene ($\hbar = 1$) 
\begin{align}
H = \int_{\vec{x}} \Psi^{\dagger}(\vec{x}) \hat{\mathcal{H}}_D \Psi(\vec{x}) + \frac{1}{2} \int_{\vec{x}, \vec{y}} \delta n(\vec{x}) V(\vec{x} - \vec{y}) \delta n(\vec{y}) \,,
\label{eq:Hamiltonian}
\end{align}
with $\delta n(\vec{x}) = \Psi^{\dagger}(\vec{x}) \Psi(\vec{x}) - \bar{n}(x)$. The fermionic field operators $\Psi$ and $\Psi^{\dagger}$ are, in fact, eight-component spinors $\Psi \equiv \begin{pmatrix} \Psi_{\uparrow} & \Psi_{\downarrow}	
\end{pmatrix}^{\intercal}$, with $\Psi_{\sigma} \equiv \begin{pmatrix} \psi_{A K_+} & \psi_{B K_+} & \psi_{B K_-} & \psi_{A K_-} \end{pmatrix}_{\sigma}^{\intercal}$. The indices $A/B$, $K_{\pm}$, and $\uparrow, \downarrow$ represent sublattice, valley and spin degrees of freedom, respectively.

The first term - the Dirac part of the Hamiltonian - describes the dynamics of the four flavours of Dirac electrons $\alpha = (K_+\!\uparrow, K_-\!\uparrow, K_+\!\downarrow, K_-\!\downarrow)$. Within the basis chosen above, the single-particle Hamiltonian $\hat{\mathcal{H}}_D$ assumes a diagonal form in flavour space
\begin{equation}
\hat{\mathcal{H}}_D = \textrm{diag} \left( \mathcal{H}_{D, K_+ \uparrow}, \mathcal{H}_{D, K_- \uparrow}, \mathcal{H}_{D, K_+ \downarrow}, \mathcal{H}_{D, K_- \downarrow} \right) \,,
\label{eq:SingleParticleHamiltonian}
\end{equation}
where the Hamiltonian for each individual flavour reads
\begin{equation}
\mathcal{H}_{D, \alpha} = - \kappa_{\alpha} i v_F \vec{\sigma} \cdot \vec{\nabla} \,.
\label{eq:SingleParticleHamiltonianPerFlavour}
\end{equation}
Here, $\kappa_{\alpha} = \pm 1$ distinguishes between the two valleys $K_{\pm}$, and $v_F$ is the Fermi velocity with the numerical value $v_F \approx c/300$. Note that we indicated the $4 \times 4$ matrix structure of the flavour space in Eq.~\eqref{eq:SingleParticleHamiltonian} with a hat symbol explicitly, while the $2 \times 2$ matrix structure of the sublattice space is implicit.

The second term of Eq.~\eqref{eq:Hamiltonian} describes two-particle interactions between the Dirac fermions. The interaction amplitude is given by the instantaneous, $U(4)$ symmetric Coulomb interaction
\begin{equation}
V(\vec{x} - \vec{y}) = \frac{e^2}{\epsilon |\vec{x} - \vec{y}|} \,.
\label{eq:CoulombInteraction}
\end{equation}
The term $\bar{n}(\vec{x}) = \sum_{\alpha} \bar{n}_{\alpha}(\vec{x})$ in the definition of the bosonic operator $\delta n(\vec{x})$ is a background density. In general it is space- and possibly even time-dependent, but for our purposes, however, will be constant. It acts as a counterterm, that cancels the zero momentum singularity of the bare Coulomb interaction. Furthermore, $\epsilon$ is the dielectric constant of the medium (being unity in vacuum), which describes the influence of a substrate on the bare Coulomb interaction.

\begin{figure}
	\centering
		\includegraphics[width=.45\textwidth]{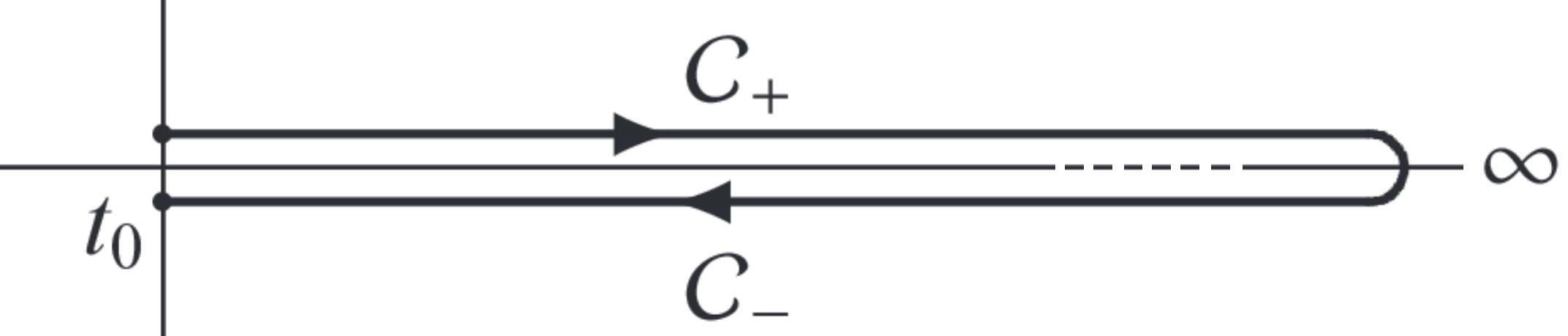}
	\caption{Schwinger-Keldysh closed time contour in the complex time plane with forward ($\mathcal{C}_+$) and backward time branch ($\mathcal{C}_-$). Here, $t_0$ is a reference time, where an initial density matrix enters the theory. Since we are only interested in the system's linear response properties close to thermal equilibrium, we send the reference time to the remote past ($t_0 = - \infty$) outright.}
	\label{fig:TimeContour}
\end{figure}

In this paper we employ the Keldysh formalism to formulate a real-time theory at finite temperature and density for the four interacting flavours of Dirac particles in graphene, that are subject to an external magnetic field and coupled to four statistical $U(1)$ gauge fields. Within the Keldysh formulation the dynamical degrees of freedom of the theory are defined on the Schwinger-Keldysh contour, which is a closed contour in the complex time plane.~\cite{ChouSuYu1985, KamenevBook} The time arguments of the field operators are elevated to contour-time and correlation functions are derived as the expectation value of their ``path ordered'' products. As shown in Fig.~\ref{fig:TimeContour}, the time contour starts at a reference time $t_0$ - at which the initial density matrix is specified - extends into the infinite future along the real axis and returns to the reference time eventually.
Here, we are mainly interested in the thermal equilibrium state in linear response to an external electromagnetic perturbation. Therefore, we send the reference time $t_0$ to the infinite past, which erases all the information about possible nontrivial initial correlations and transient regimes.~\cite{Danielewicz1984, CalzettaHu1988} As a consequence, the quantum kinetic equations are of no further concern, since they can be trivially solved by the well-known thermal distribution functions.~\cite{KamenevBook}

Before we discuss the field theoretic model in its action formulation a few remarks concerning notational conventions are in order. First, to a large extent we will work within the abstract contour-time representation, and only switch to a physical real-time representation at the end of Sec.~\ref{sec:EffectiveBosonicActionMeanFieldTheoryAndGaussianFluctuations} when we discuss Gaussian fluctuations of the bosonic effective action around the mean field solution of the fractional quantum Hall liquid. The major advantage of the contour-time representation is that it allows for a compact and concise notation, resembling the zero temperature vacuum (or ground state) theory, yet encoding the full information of thermal fluctuations.~\cite{ChouSuYu1985} Furthermore we employ a covariant notation, where upper and lower case greek letters $\mu, \nu, \lambda$ denote contra- and covariant components of a Minkowski three-vector, respectively. As usual a repeated index implies summation according to the Einstein summation convention. This summation rule is lifted if a repeated index is bracketed. (This statement will only apply for repeated flavour space indices $\alpha, \beta$, see Eq.~\eqref{eq:InverseContourTimeFlavourPropagator} for instance.) The convention for the flat Minkowski metric is chosen to be $\eta_{\mu \nu} = \textrm{diag}(1, -1, -1)$, and natural units ($\hbar = c = 1$) are used throughout the article. Lastly, space-time integrations will be denoted by
\begin{equation}
\int_{\mathcal{C}, x} \equiv \int_{\mathcal{C}} dt \int d^2r \,,
\label{eq:SpaceTimeIntegration}
\end{equation}
where $\mathcal{C}$ indicates that the time integration is performed along the Schwinger-Keldysh contour, and $x = x^{\mu} = (t, \vec{r})$ labels (contour-)time and spatial variables. After introducing these general notational conventions we proceed to describe the details of the model.

The entire physical content of the theory is summarized by the coherent state functional integral~\cite{NegeleOrlandBook, ChouSuYu1985, KamenevBook}
\begin{equation}
Z[e A_{\mu} + \mathcal{A}_{\mu}^{\alpha}] = \!\int\! \mathcal{D}\psi \mathcal{D}\psi^{\dagger} \mathcal{D}a \, e^{i S[\psi, e A_{\mu} + \mathcal{A}_{\mu}^{\alpha}, a_{\mu}^{\alpha}]} \,,
\label{eq:DefinitionContourTimeGeneratingFunctional}
\end{equation}
which is a generating functional of correlation functions. The action $S$ in the exponential can be written as a sum of three terms
\begin{align}
S[\psi, e A_{\mu} + \mathcal{A}_{\mu}^{\alpha}, a_{\mu}^{\alpha}] &= S_D[\psi, e A_{\mu} + \mathcal{A}_{\mu}^{\alpha} + a_{\mu}^{\alpha}] + S_{Coul}[\psi] \nonumber \\
&+ S_{CS}[a_{\mu}^{\alpha}] \,,
\label{eq:Action}
\end{align}
where the first two terms, involving the fermionic fields, are readily obtained from the Heisenberg picture Hamiltonian $H(t)$ by the definition
\begin{equation}
S_D[\psi] + S_{Coul}[\psi] = \int_{\mathcal{C}, t} \left( \int_{\vec{r}} \Psi^{\dagger}(x) i \partial_t \Psi(x) - H(t) \right) \,.
\label{eq:MappingHamiltonianToAction}
\end{equation}

The bosonic fields within the Dirac part of the action, $A_{\mu}$, $\mathcal{A}_{\mu}^{\alpha}$ and $a_{\mu}^{\alpha}$, are introduced via the minimal coupling prescription. They represent an external electromagnetic potential, local two-particle source fields, and the statistical gauge fields respectively. The source fields will later be used to generate the desired correlation functions. The Dirac action can be written compactly as a quadratic form of an eight-component Grassmann spinor~$\Psi$~\cite{GusyninSharapovCarbotte2007, GorbarGusynin2002}
\begin{equation}
S_D[\psi, e A_{\mu} + \mathcal{A}_{\mu}^{\alpha} + a_{\mu}^{\alpha}] = \int_{\mathcal{C}, xy} \Psi^{\dagger}(x) \hat{G}_0^{-1}(x, y) \Psi(y) \,.
\label{eq:ActionFermionicQuadraticPart}
\end{equation}
The matrix $\hat{G}_0^{-1}$ is the inverse contour-time propagator, which inherits the flavour diagonal structure from the single-particle Hamiltonian~\eqref{eq:SingleParticleHamiltonian}
\begin{equation}
\hat{G}_0^{-1} = \textrm{diag} \left( G_{0, K_+ \uparrow}^{-1}, G_{0, K_- \uparrow}^{-1}, G_{0, K_+ \downarrow}^{-1}, G_{0, K_- \downarrow}^{-1} \right) \,.
\label{eq:InverseFreeContourTimePropagatorFlavourSpaceStructure}
\end{equation}
According to Eq.~\eqref{eq:SingleParticleHamiltonianPerFlavour} the dynamics of each flavour is governed by the pseudorelativistic, massless Weyl operator
\begin{align}
G_{0, \alpha}^{-1}(x, y) = \delta_{\mathcal{C}}(x - y) \left( i \sigma_{(\alpha)}^{\mu} \mathcal{D}_{\mu}^{(\alpha)} + \mu_{\alpha} \right) \,.
\label{eq:InverseContourTimeFlavourPropagator}
\end{align}
Here, $\delta_{\mathcal{C}}(x - y) = \delta_{\mathcal{C}}(x_0 - y_0) \delta(\vec{x} - \vec{y})$ involves the contour-time delta function~\cite{ChouSuYu1985} and  $\sigma_{\alpha}^{\mu} \equiv (\sigma_0, \kappa_{\alpha} v_F \sigma_1, \kappa_{\alpha} v_F \sigma_2)$ is a three-vector of Pauli matrices, acting in sublattice space. 
The gauge covariant derivative
\begin{align}
\mathcal{D}_{\mu}^{\alpha} &= \partial_{\mu} + ie A_{\mu}(x^{\mu}) + i \mathcal{A}_{\mu}^{\alpha}(x^{\mu}) + i a_{\mu}^{\alpha}(x^{\mu}) \,,
\label{eq:CovariantDerivative}
\end{align}
contains the aforementioned covariant vector potentials $A_{\mu}, \mathcal{A}_{\mu}^{\alpha}$ and $a_{\mu}^{\alpha}$.  For the external potential $A_{\mu}$ we choose the Landau gauge, $A_{\mu}(x^{\mu}) = (0, Bx^2, 0) = (0, By, 0)$, to describe a uniform and static magnetic field $B$ perpendicular to the graphene plane. Note that it does not depend on the flavour index $\alpha$, so that all flavours universally couple to the same field. The source fields $\mathcal{A}_{\mu}^{\alpha}$ and the statistical gauge fields $a_{\mu}^{\alpha}$, on the other hand, do carry a flavour index and, thus, couple to each fermionic flavour individually. Such a coupling breaks the global $U(4)$ symmetry of the theory without Chern-Simons fields down to a local $U(1)^{\otimes 4}$ symmetry. Finally, we introduced a flavour dependent chemical potential $\mu_{\alpha}$, allowing for independent doping of the individual flavours. Physically this flavour dependence may be thought of as originating form a generalized Zeeman term.~\cite{Goerbig2011}

The Coulomb interaction part requires no further discussion as it is directly obtained from the interaction part of the Hamiltonian~\eqref{eq:Hamiltonian}
\begin{equation}
S_{Coul}[\psi] = - \frac{1}{2} \int_{\mathcal{C}, xy} \delta n(x) V(x - y) \delta n(y) \,,
\label{eq:CoulombInteractionAction}
\end{equation}
with $V(x - y) = V(\vec{x} - \vec{y}) \delta_{\mathcal{C}}(x_0 - y_0)$.

The third term in the action~\eqref{eq:Action} is the kinetic term for the four statistical gauge fields, which is given by a generalized Chern-Simons action~\cite{LopezFradkin1991, LopezFradkin1993, LopezFradkin1995, FradkinBook}
\begin{equation}
S_{CS}[a_{\mu}^{\alpha}] = \frac{1}{2} (\hat{\mathcal{K}}^{-1})_{\alpha \beta} \int_{\mathcal{C}, x} \varepsilon^{\mu \nu \lambda} a_{\mu}^{\alpha}(x) \partial_{\nu} a_{\lambda}^{\beta}(x) \,.
\label{eq:ChernSimonsAction}
\end{equation}
Herein $\varepsilon^{\mu \nu \lambda}$ is the total antisymmetric Levi-Civita tensor (we use the convention $\varepsilon^{012} = 1$), and $\hat{\mathcal{K}}$ is a regular, \textit{i.e.} invertible, symmetric $4 \times 4$ matrix
\begin{equation}
\hat{\mathcal{K}} = 2 \pi \begin{pmatrix} 2 k_1 & m_1 & n_1 & n_2 \\ m_1 & 2 k_2 & n_3 & n_4 \\ n_1 & n_3 & 2 k_3 & m_2 \\ n_2 & n_4 & m_2 & 2 k_4 \end{pmatrix} \,,
\label{eq:DefinitionKMatrix}
\end{equation}
with integers $k_i, m_i, n_i$. For those configurations of integers where $\hat{\mathcal{K}}$ happens to be singular Eq.~\eqref{eq:ChernSimonsAction} needs to be regularized. This may be achieved by adding a diagonal matrix $\hat{\mathcal{R}} = 2\pi \textrm{diag}(+i \eta, -i \eta, +i \eta, -i \eta)$ to Eq.~\eqref{eq:DefinitionKMatrix}, where $\eta$ is an infinitesimal (the signs therein are purely conventional). The physical meaning of the $\mathcal{K}$-matrix is to attach statistical magnetic flux to the fermions. This feature will become more clear in the next section when we discuss the stationary phase approximation.

The theory we described above possesses a \textit{local} $U(1)^{\otimes 4}$ symmetry, in comparison to the symmetry of the original model of interacting electrons in graphene, being a \textit{global} $U(4)$ flavour symmetry ($U(2) \times U(2)$ respectively, if one takes into account a Zeeman term~\cite{GusyninSharapovCarbotte2007}). It has to be emphasized that the symmetry is broken explicitly by considering the flavour dependent chemical potential in addition to the $U(1)^{\otimes 4}$ symmetric gauge field coupling. As pointed out by the authors of Ref.~[\onlinecite{LopezFradkin1995}], who studied the FQHE for nonrelativistic fermions in bilayers, as well as $SU(2)$ symmetric monolayers, the original symmetry $U(4)$ may only be generated dynamically (once the flavour dependence of the chemical potential is neglected~\cite{LopezFradkin1995}). Therefore, it is expected that some of the fractional quantum Hall states we obtain in this work - after certain necessary approximations have been made - may not be realized in the exact theory, as they could be destabilized by higher order fluctuations. A manifestly $U(4)$ (respectively $U(2) \times U(2)$) symmetric theory, on the other hand, could be constructed by considering an appropriate nonabelian generalization of Eq.~\eqref{eq:ChernSimonsAction}, with a corresponding set of nonabelian statistical gauge fields, coupling gauge covariantly to the fermions.~\cite{LopezFradkin1995} Clearly, such a nonabelian gauge theory is in many aspects significantly more complex than the abelian theory of the present article and we leave its construction and analysis for future work.

As a final remark we want to stress that the partition function \eqref{eq:DefinitionContourTimeGeneratingFunctional} as it stands is not well-defined. Since the Chern-Simons fields $a_{\mu}^{\alpha}$ are gauge fields, the functional integral contains an infinite summation over all, physically equivalent orbits of pure gauge, leading to a strong divergence. In order to extract physically meaningful information from the partition function, the gauge equivalent orbits have to be removed, such that each gauge field configuration in the functional integral uniquely corresponds to a physical field configuration. To this end we employ the well-known Fadeev-Popov procedure,~\cite{PeskinSchroederBook} but we postpone the details of the discussion to Sec.~\ref{sec:ElectromagneticResponseTensorAndHallConductance}. For now we work with Eqs.~\eqref{eq:DefinitionContourTimeGeneratingFunctional} and \eqref{eq:Action} as they are, but keep in mind that they need to be modified.

\section{Effective Bosonic Action, Mean Field Theory and Gaussian Fluctuations}
\label{sec:EffectiveBosonicActionMeanFieldTheoryAndGaussianFluctuations}
In this section we derive an exact expression for the effective action of the gauge fields $a_{\mu}^{\alpha}$, following Ref.~[\onlinecite{LopezFradkin1991}]. Subsequently, the nonpolynomial action we obtain will be expanded to second order in the fluctuations around its mean-field solution, resulting in an exactly solvable Gaussian model. The quadratic action will be stated in its real-time form in Keldysh basis. 

Due to the Coulomb interaction being quartic in the fermionic fields, an integration of these microscopic degrees of freedom is not readily possible. For this reason we rewrite the problematic interaction term by means of a Hubbard-Stratonovich transformation in the density-density channel,~\cite{KamenevBook} which introduces an auxiliary boson~$\phi$
\begin{equation}
e^{i S_{int}[\psi]} = \int \mathcal{D}\phi \, e^{i S_{HS}[\phi] + i S_{int}[\psi, \phi]} \,.
\label{eq:HubbardStratonovichTransformationInteraction}
\end{equation}
The quadratic action of the Hubbard-Stratonovich boson is given by
\begin{equation}
S_{HS}[\phi] = \frac{1}{2} \int_{\mathcal{C}, xy} \phi(x) V^{-1}(x - y) \phi(y) \,,
\label{eq:BosonActionQuadraticPart}
\end{equation}
with the inverse Coulomb interaction $V^{-1}$, which, of course, has to be understood in the distributional sense. The second term contains a trilinear Yukawa-type interaction and a linear term, describing the interaction of the auxiliary boson with the background density $\bar{n}$
\begin{equation}
S_{int}[\psi, \phi] = - \int_{\mathcal{C}, x} \phi(x) \left( \Psi^{\dagger}(x) \Psi(x) - \bar{n}(x) \right) \,.
\label{eq:ActionFermiBoseCoupling}
\end{equation}
Note that the fluctuating Bose field $\phi$ in the Yukawa interaction appears on the same footing as the zero component of the external gauge potential $A_{\mu}$, coupling to all flavours identically, see Eqs.~\eqref{eq:ActionFermionicQuadraticPart}-\eqref{eq:CovariantDerivative}. As a consequence of the above manipulation the Grassmann fields $\psi$ appear only quadratically, such that the fermionic integral can be performed exactly. Our intermediate result for the effective action now only contains bosonic degrees of freedom
\begin{align}
S_{\textrm{eff}}'[e A_{\mu} + \mathcal{A}_{\mu}^{\alpha}, a_{\mu}^{\alpha}, \phi] = 
&-i \textrm{tr ln} \, \hat{G}_0^{-1}[e A_{\mu} + \mathcal{A}_{\mu}^{\alpha} + a_{\mu}^{\alpha} + \phi \delta_{0 \mu}] \nonumber \\
&+ S_{HS}[\phi] + \phi \bar{n} + S_{CS}[a_{\mu}^{\alpha}] \,.
\label{eq:ExactEffectiveAction1}
\end{align}

Remarkably, the Hubbard-Stratonovich boson $\phi$ can be integrated exactly after shifting the statistical gauge fields as follows: $a_{\mu}^{\alpha} \rightarrow a_{\mu}^{\alpha} - \phi \delta_{0 \mu}$.~\cite{LopezFradkin1991, LopezFradkin1995} The result is the desired effective action of the Chern-Simons gauge fields in the presence of the two-particle source fields $\mathcal{A}_{\mu}^{\alpha}$
%
%
\begin{align}
S_{\textrm{eff}}[e A_{\mu} + \mathcal{A}_{\mu}^{\alpha}, a_{\mu}^{\alpha}] = &-i \textrm{tr ln} \, \hat{G}_0^{-1}[e A_{\mu} + \mathcal{A}_{\mu}^{\alpha} + a_{\mu}^{\alpha}] \nonumber \\
&+ S_V[a_{\mu}^{\alpha}] + S_{CS}[a_{\mu}^{\alpha}] \,.
\label{eq:ExactEffectiveAction2}
\end{align}
The term $S_V[a_{\mu}^{\alpha}]$ is a quadratic functional of the gauge fields, that is generated by the $\phi$-integration
\begin{widetext}

\begin{equation}
S_V[a_{\mu}^{\alpha}] = - \frac{1}{2} \int_{\mathcal{C}, xy} \Big((\hat{\mathcal{K}}^{-1})_{\alpha_1 \beta_1} \varepsilon^{0 \mu_1 \nu_1} \partial_{\mu_1} a_{\nu_1}^{\beta_1} - \bar{n}_{\alpha_1} \Big)(x) V^{\alpha_1 \alpha_2}(x - y) \Big((\hat{\mathcal{K}}^{-1})_{\alpha_2 \beta_2} \varepsilon^{0 \mu_2 \nu_2} \partial_{\mu_2} a_{\nu_2}^{\beta_2} - \bar{n}_{\alpha_2} \Big)(y) \,.
\label{eq:NewInteraction}
\end{equation}
Here we have defined $V^{\alpha \beta}(x - y) \equiv V(x - y)$, where the additional flavour-space indices keep track of the correct summation. Note that Eq.~\eqref{eq:NewInteraction} is nothing but the Coulomb interaction term~\eqref{eq:CoulombInteractionAction}, in which the density of flavour $\alpha$, $\Psi_{(\alpha)}^{\dagger}(x) \Psi_{(\alpha)}(x)$, is substituted by $(\hat{\mathcal{K}}^{-1})_{\alpha \beta} \varepsilon^{0 \mu \nu} \partial_{\mu} a_{\nu}^{\beta}(x)$. 
%
%
%
In the above derivation no approximations were involved. Yet, due to the nonpolynomial tracelog term, the residual functional integral over the gauge fields cannot be performed exactly. A common strategy to deal with this problem, which we adopt here as well, is to find the field configuration in which the effective action becomes stationary and, subsequently, expand in powers of fluctuations around the mean. 

The variation of Eq.~\eqref{eq:ExactEffectiveAction2} in the absence of two-particle sources $\mathcal{A}_{\mu}^{\alpha}$ yields

\begin{align}
\frac{\delta S_{\textrm{eff}}}{\delta a_{\mu}^{\alpha}(z)} = -& j_{\alpha}^{\mu}(z) + (\hat{\mathcal{K}}^{-1})_{\alpha \beta} \varepsilon^{\mu \nu \lambda} \partial_{\nu} a_{\lambda}^{\beta}(z) \nonumber \\
-& \int_{\mathcal{C}, xy} \Big((\hat{\mathcal{K}}^{-1})_{\alpha_1 \beta_1} \varepsilon^{0 \mu_1 \nu_1} \partial_{\mu_1} \delta_{\nu_1}^{\mu} \delta_{\alpha}^{\beta_1} \delta_{\mathcal{C}}(x - z) \Big) V^{\alpha_1 \alpha_2}(x - y) \Big((\hat{\mathcal{K}}^{-1})_{\alpha_2 \beta_2} \varepsilon^{0 \mu_2 \nu_2} \partial_{\mu_2} a_{\nu_2}^{\beta_2} - \bar{n}_{\alpha_2} \Big)(y) \,,
\label{eq:MeanFieldEquations}
\end{align}

\end{widetext}
Here $j_{\alpha}^{\mu}$ is the particle 3-current density per flavour $\alpha$ in the presence of an external gauge potential $A_{\mu}$ and the Chern-Simons fields $a_{\mu}^{\alpha}$
\begin{equation}
j_{\alpha}^{\mu}(x) = -i \frac{\delta}{\delta a_{\mu}^{\alpha}(x)} \textrm{tr ln} \, \hat{G}_0^{-1}[e A_{\mu} + a_{\mu}^{\alpha}] \,.
\label{eq:DefinitionParticle3CurrentDensity}
\end{equation}
We have to stress at this point that Eqs.~\eqref{eq:MeanFieldEquations} and \eqref{eq:DefinitionParticle3CurrentDensity} have to be treated with special care as they demand a proper regularization. First, in the infinite system the particle current is not well defined, since its $\mu = 0$ component - being the particle density - diverges. This fact is a direct consequence of the Dirac approximation of the tight-binding graphene spectrum. Another issue is related to the fact that the definition of the particle current involves the average of a (contour-)time ordered product of two fermionic fields evaluated at the same time. However, these problems are immediately resolved once the theory is mapped to the physical real-time representation in Keldysh-basis. Hereto, one splits the Schwinger-Keldysh contour into forward and backward branch and defines a doubled set of fields, $\Psi_{\pm}$ and $(a_{\pm})_{\mu}^{\alpha}$, which are associated to the respective branch.~\cite{ChouSuYu1985, KamenevBook} In a next step, one performs a rotation from $\pm$-basis to Keldysh-basis by defining ``classical'' and ``quantum'' fields, indexed by $c$ and $q$ respectively, as symmetric and antisymmetric linear combinations of the $\pm$-fields.~\cite{ChouSuYu1985, KamenevBook} The net result is that the derivative in Eq.~\eqref{eq:MeanFieldEquations} is performed with respect to the quantum components of the gauge fields, the particle 3-current densities are replaced by the well-defined charge carrier 3-current densities $\bar{j}_{\mu}^{\alpha}(x)$, see Eq.~\eqref{eq:ChargeCarrierCurrent}, and the gauge fields on the right hand side are replaced by their classical components.~\cite{Note2}

The requirement of a vanishing first variation defines the mean-field equations for the Chern-Simons fields. As pointed out by the authors of Refs.~[\onlinecite{LopezFradkin1991}] and [\onlinecite{LopezFradkin1995}], these mean-field equations allow for several physically different scenarios such as Wigner crystals and solitonic field configurations. Following Refs.~[\onlinecite{LopezFradkin1991}] and [\onlinecite{LopezFradkin1995}], we here concentrate on those solutions, which lead to a vanishing charge carrier current and a uniform and time independent charge carrier density $\bar{n}_{\alpha}$, describing a quantum Hall liquid. In that case, Eq.~\eqref{eq:MeanFieldEquations} reduces to the relation
\begin{equation}
\bar{n}_{\alpha} = (\hat{\mathcal{K}}^{-1})_{\alpha \beta} \varepsilon^{0 \mu \nu} \partial_{\mu} \bar{a}_{\nu}^{\beta} = -e (\hat{\mathcal{K}}^{-1})_{\alpha \beta} b^{\beta} \,.
\label{eq:SolutionMeanFieldEquations}
\end{equation}
Here, the second equality defines the (uniform) Chern-Simons magnetic field $b^{\alpha}$, experienced by the flavour $\alpha$ charge carriers, in terms of the expectation value of the Chern-Simons fields $\bar{a}_{\mu}^{\alpha} \equiv \langle (a_c)_{\mu}^{\alpha} \rangle$. Inverting the above relation yields the statistical magnetic fields $b^{\alpha}$ as functions of the densities $\bar{n}_{\alpha}$
\begin{equation}
b^{\alpha} = - \frac{1}{e} \mathcal{K}^{\alpha \beta} \bar{n}_{\beta} \,.
\label{eq:SolutionMeanFieldEquations1}
\end{equation}
Writing the mean-field equation in this form reveals the physical meaning of the $\mathcal{K}$-matrix, as it defines the precise flux-attachment procedure of the multicomponent quantum Hall system. Each flavour $\beta$ of charge carriers, contributes to the statistical magnetic field for the flavour $\alpha$ with a magnetic flux $\mathcal{K}^{\alpha (\beta)} \bar{n}_{(\beta)}$. Hence, the component $\mathcal{K}^{\alpha \beta}$ represents the contribution to the statistical flux per flavour $\beta$ as seen by flavour $\alpha$. Thus, Eq.~\eqref{eq:SolutionMeanFieldEquations1} may be interpreted as a ``flux-binding'' relation, which transforms ordinary Dirac fermions into ``composite Dirac fermions''. Furthermore, it is important to notice that Eq.~\eqref{eq:SolutionMeanFieldEquations1} is well-defined even for singular $\mathcal{K}$-matrices, in contrast to Eq.~\eqref{eq:SolutionMeanFieldEquations}. Such singular configurations should not be discarded, however, as the following discussion shows. Consider for example the special case, where all components of $\hat{\mathcal{K}}$ equal the same constant $2k$. In that case the four equations~\eqref{eq:SolutionMeanFieldEquations1} reduce to a single one, yielding a unique statistical field $b$ associated to the density of charge carriers $\bar{n} = \sum_{\alpha} \bar{n}_{\alpha}$. This scenario corresponds to a Chern-Simons theory, where only a single dynamical gauge field, $a_{\mu} = \sum_{\alpha} a_{\mu}^{\alpha}$, is present, that couples to the different flavours identically.~\cite{ModakEtal2011} The other three eigenvectors one obtains by diagonalizing Eq.~\eqref{eq:DefinitionKMatrix} span a triply degenerate subspace with eigenvalue zero, and thus decouple. Likewise, for other singular $\mathcal{K}$-matrix configurations one would obtain a theory with only two or three dynamical gauge fields and a correspondingly reduced parameter space. (In the extreme case where $\hat{\mathcal{K}}$ is identically zero, all gauge fields would decouple and no flux-binding could occur, which leads back to the integer quantum Hall regime.) With this physical picture in mind we now continue our discussion.

By virtue of the gauge covariant derivative \eqref{eq:CovariantDerivative} each one of the statistical magnetic fields \eqref{eq:SolutionMeanFieldEquations1} adds to the external magnetic field $B$ individually, resulting in a flavour dependent effective magnetic field~\cite{Note3}
%
\begin{equation}
B_{\textrm{eff}}^{\alpha} = B + b^{\alpha} = B - \frac{1}{e} \mathcal{K}^{\alpha \beta} \bar{n}_{\beta} \,.
\label{eq:SelfConsistentDefinitionEffectiveMagneticField}
\end{equation}
It is this effective magnetic field, rather than the external field $B$ alone, which enters the fermionic propagators, such that Eqs.~\eqref{eq:SolutionMeanFieldEquations1} and~\eqref{eq:SelfConsistentDefinitionEffectiveMagneticField}, in fact, represent self-consistency equations. A straightforward calculation of the free propagator for Dirac fermions moving in the effective magnetic field $B_{\textrm{eff}}^{\alpha}$ yields the charge carrier density for the flavour $\alpha$ as a function of the chemical potential $\mu_{\alpha}$, the effective magnetic field $B_{\textrm{eff}}^{\alpha}$ and temperature $T$, (see App.~\ref{sec:FermionPropagatorInExternalMagneticField} for details)
\begin{equation}
\bar{n}_{\alpha}(\mu_{\alpha}, B_{\textrm{eff}}^{\alpha}, T) = \frac{1}{2 \pi \ell_{(\alpha)}^2} \nu_{(\alpha)}(\mu_{\alpha}, B_{\textrm{eff}}^{\alpha}, T) \,.
\label{eq:ChargeCarrierDensity}
\end{equation}
%
%
Here, we have introduced the magnetic length $\ell_{\alpha} = 1/\sqrt{|e B_{\textrm{eff}}^{\alpha}|}$ and the filling fraction per flavour~\cite{GorbarGusynin2002, GusyninSharapov2005}
\begin{equation}
\nu_{\alpha} = \frac{1}{2} \left( \textrm{tanh} \frac{\mu_{\alpha}}{2T} + \sum_{n=1}^{\infty} \sum_{\lambda = \pm} \textrm{tanh} \frac{\lambda \sqrt{n} \omega_{\textrm{c}}^{\alpha} + \mu_{\alpha}}{2T} \right) \,,
\label{eq:FlavourFillingFraction}
\end{equation}
where $\omega_{\textrm{c}}^{\alpha} = \sqrt{2} v_F/\ell_{\alpha}$ denotes the pseudo-relativistic cyclotron frequency. The charge carrier density as a function of an effective magnetic field $B_{\textrm{eff}}^{\alpha}$ at constant chemical potential $\mu_{\alpha}$ and the filling fraction as a function of the chemical potential at constant field, are shown in Fig.~\ref{fig:HallPlots}. At large magnetic fields and low temperatures the filling fraction shows the typical plateau structure that is characteristic for the (anomalous) integer quantum Hall effect. This issue will be discussed in more detail at the end of this section, once we have obtained the Gaussian approximation to the exact action~\eqref{eq:ExactEffectiveAction2}. 

From the above mean field equation one may calculate the possible fractional fillings at which the spectrum is gapped, leading to a plateau structure for the ``interacting Hall conductivity'', which is the hallmark of the fractional quantum Hall effect. However, we prefer to extract the filling fractions directly from the interacting Hall conductivity, which will be derived in the next section. To continue we only need to know that the mean field equation has a nontrivial solution, which depends on the $\mathcal{K}$-matrix, the external magnetic field, temperature and chemical potential.
Furthermore, observe that the effective magnetic field is invariant upon changing the sign of $\mathcal{K}_{\alpha \beta}$ and $\bar{n}_{\alpha}$ simultaneously. This is a first hint, how to construct manifestly particle-hole symmetric filling fractions in the presence of a Chern-Simons term.

The stationary field configuration we found above serves as a reference point around which one should expand the effective action~\eqref{eq:ExactEffectiveAction2} in powers of field fluctuations. To this end one writes $a_{\mu}^{\alpha} = \bar{a}_{\mu}^{\alpha} + \Delta a_{\mu}^{\alpha}$ and expands the effective action to the desired order in the fluctuation $\Delta a_{\mu}^{\alpha}$ and the source $\mathcal{A}_{\mu}^{\alpha}$. As mentioned before, here we are only interested in an expansion up to second order. Terms which are linear in the fluctuation vanish, since the effective action is evaluated at the saddle point, whereas linear source terms do not vanish. However, since the latter only couple to the above mean field $3$-currents, their contribution is not interesting for the further analysis and will be omitted. We state the result in the physical real-time representation in Keldysh basis
\begin{widetext}

\begin{figure}
	\centering
		\includegraphics[width=0.32\textwidth]{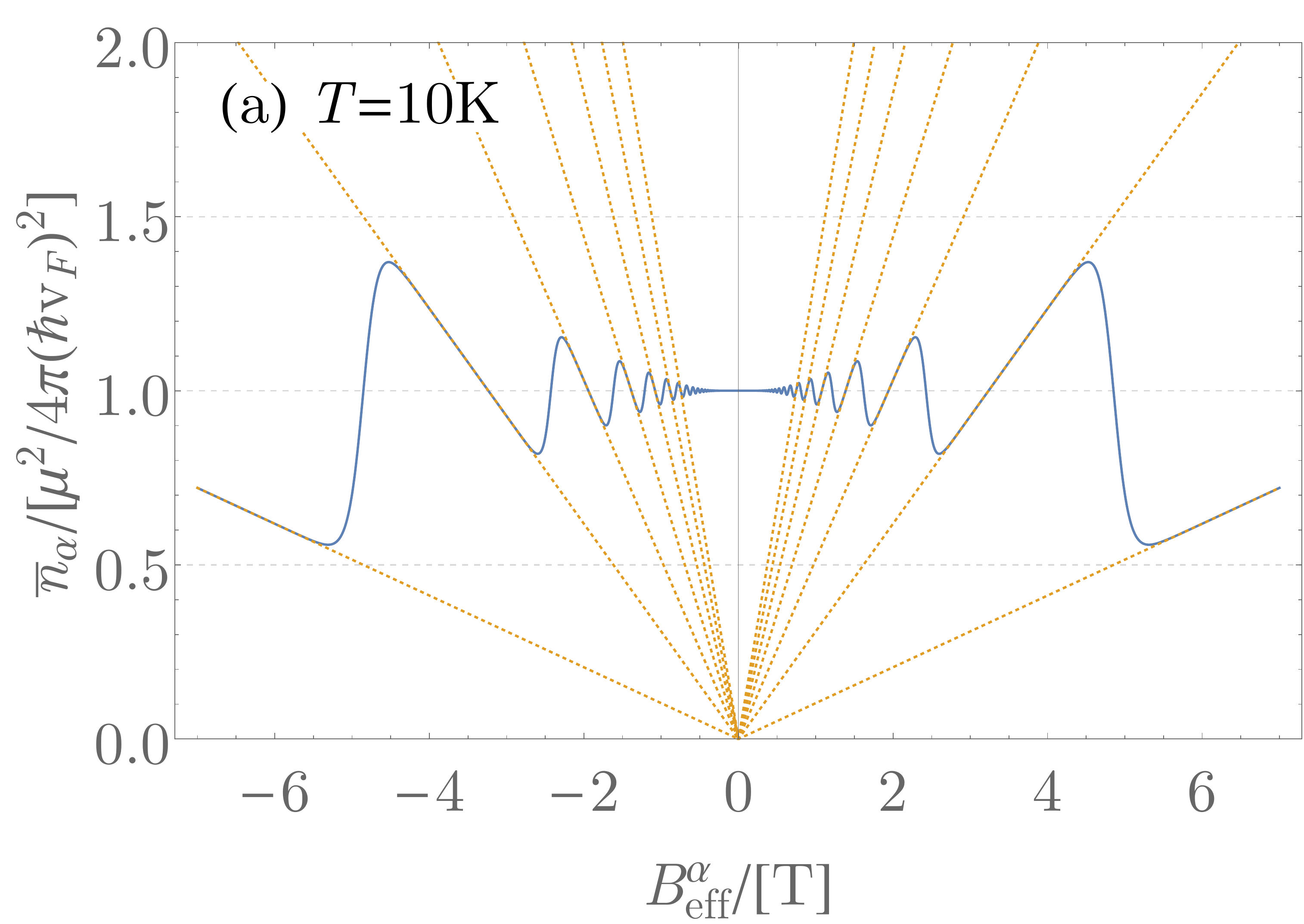}
		\includegraphics[width=0.32\textwidth]{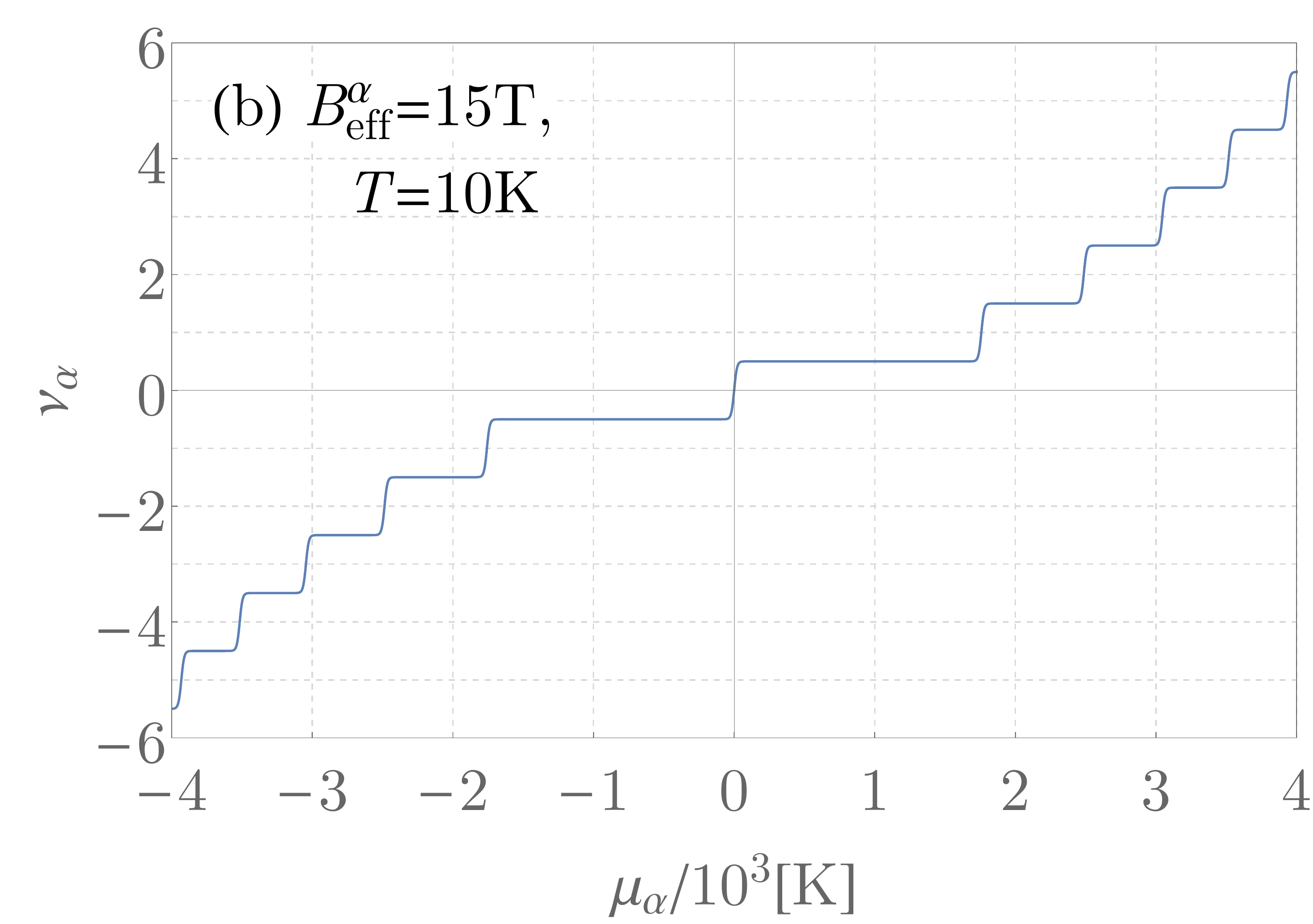}
				\includegraphics[width=0.32\textwidth]{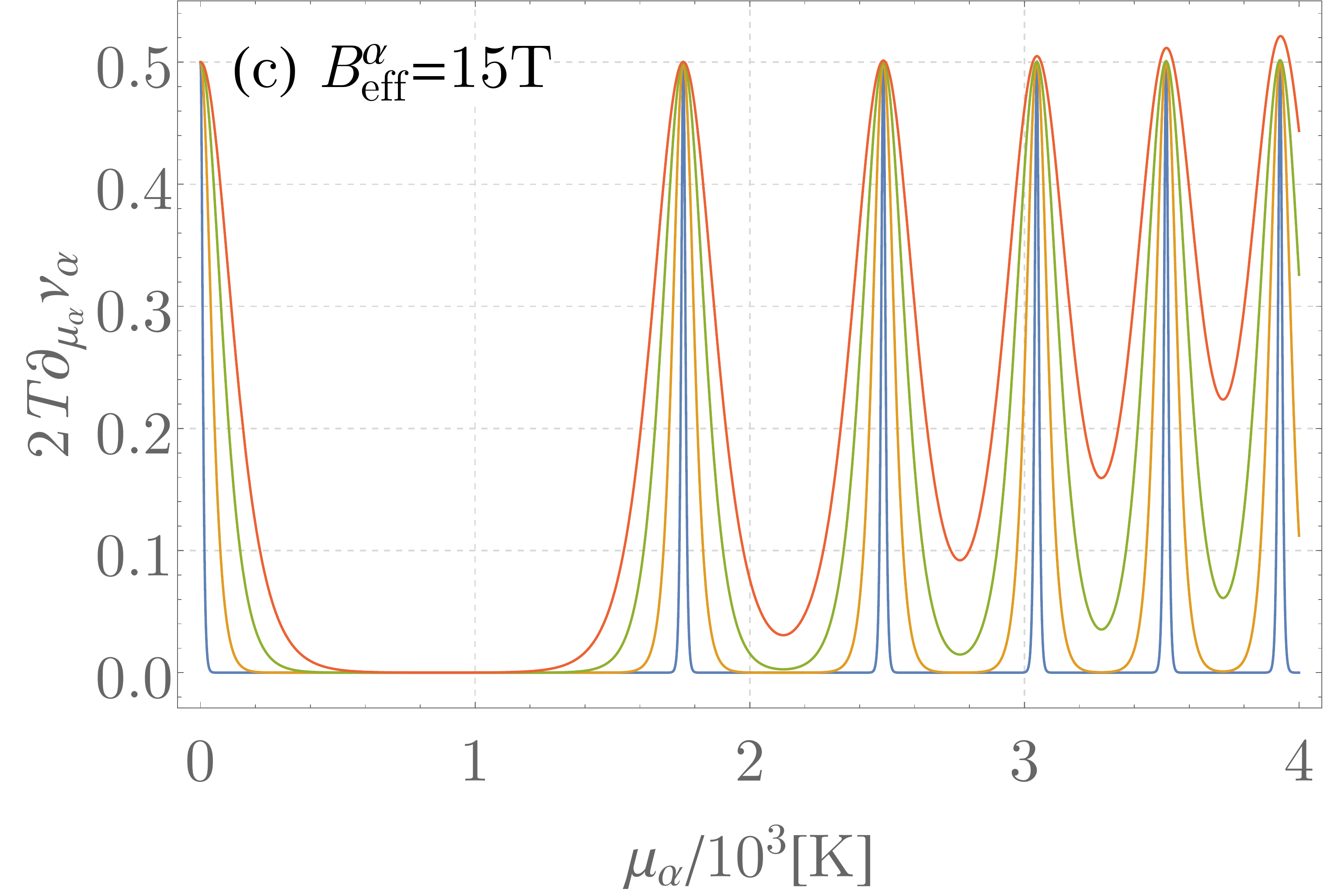}
	\caption{(Color online) (a) Charge carrier density at constant chemical potential as a function of the effective magnetic field $B_{\textrm{eff}}^{\alpha}$ at $T = 10K$ (blue). The straight dotted lines (orange) indicate the first few Landau levels. At vanishing magnetic field the charge carrier density scales quadratically with the chemical potential.~\cite{GusyninSharapov2005, GusyninSharapovCarbotte2007} (Without loss of generality the sign of the chemical is assumed to be positive). Upon increasing (the absolute value of) the effective magnetic field $B_{\textrm{eff}}^{\alpha}$ - while keeping the chemical potential fixed - the carrier density shows oscillations in the regime $\omega_{\textrm{c}}^{\alpha} < \mu_{\alpha}$, whereas for $\omega_{\textrm{c}}^{\alpha} \gg \mu_{\alpha}$ it grows linearly as a function of the magnetic field. This behaviour is readily explained by the formation of Landau levels, and the dependence of their degeneracy and relative energetic separation on the magnetic field. (b) Filling fraction per flavour $\nu_{\alpha}$ at constant effective magnetic field $B_{\textrm{eff}}^{\alpha} = 15 \textrm{T}$ and temperature $T = 10 \textrm{K}$ as a function of the chemical potential. The plateaus occur at half-integer filling fractions $\nu_{\alpha} = \pm (n_{\alpha} + 1/2)$. The transitions between the plateaus are smeared out due to temperature. (c) Derivative of $\nu_{\alpha}$ with respect to the chemical potential as a measure for temperature-induced Landau level broadening at $B_{\textrm{eff}}^{\alpha} = 15 \textrm{T}$ for the temperatures $T = 5 \textrm{K}, 25 \textrm{K}, 50 \textrm{K}, 75 \textrm{K}$. Increasing the temperature clearly leads to a broadening of the discrete energy levels. Since the relativistic Landau levels are not equidistant in energy space, the level broadening causes the Landau levels to overlap significantly away from the charge neutrality point. The Landau level located directly at the charge neutrality point, however, remains well-defined up to rather large temperatures.}
	\label{fig:HallPlots}
\end{figure}

\begin{align}
S_{\textrm{eff}}[\boldsymbol{\mathcal{A}}_{\mu}^{\alpha}, \boldsymbol{\Delta a}_{\mu}^{\alpha}] &= \int_{xy} \begin{pmatrix} (\Delta a_c)_{\mu}^{\alpha} +	(\mathcal{A}_c)_{\mu}^{\alpha} & (\Delta a_q)_{\mu}^{\alpha} +	(\mathcal{A}_q)_{\mu}^{\alpha} \end{pmatrix}(x) \begin{pmatrix} 0 & (\Pi^A)_{\alpha \beta}^{\mu \nu} \\ (\Pi^R)_{\alpha \beta}^{\mu \nu} & (\Pi^K)_{\alpha \beta}^{\mu \nu} \end{pmatrix}(x, y) \begin{pmatrix} (\Delta a_c)_{\nu}^{\beta} +	(\mathcal{A}_c)_{\nu}^{\beta} \\ (\Delta a_q)_{\nu}^{\beta} +	(\mathcal{A}_q)_{\nu}^{\beta} \end{pmatrix}(y) \nonumber \\
&+ \int_{xy} \begin{pmatrix} (\Delta a_c)_{\mu}^{\alpha} & (\Delta a_q)_{\mu}^{\alpha} \end{pmatrix}(x) \begin{pmatrix} 0 & (C^A)_{\alpha \beta}^{\mu \nu} \\ (C^R)_{\alpha \beta}^{\mu \nu} & (C^K)_{\alpha \beta}^{\mu \nu} \end{pmatrix}(x, y) \begin{pmatrix} (\Delta a_c)_{\nu}^{\beta} \\ (\Delta a_q)_{\nu}^{\beta} \end{pmatrix}(y) \nonumber \\
&\equiv \int_{xy} \left[ \Big(\boldsymbol{\Delta a}_{\mu}^{\alpha} + \boldsymbol{\mathcal{A}}_{\mu}^{\alpha} \Big)^{\intercal}(x) \boldsymbol{\Pi}_{\alpha \beta}^{\mu \nu}(x, y) \Big(\boldsymbol{\Delta a}_{\nu}^{\beta} + \boldsymbol{\mathcal{A}}_{\nu}^{\beta} \Big)(y) +  \boldsymbol{\Delta a}_{\mu}^{\alpha}(x) \boldsymbol{C}_{\alpha \beta}^{\mu \nu}(x, y) \boldsymbol{\Delta a}_{\nu}^{\beta}(y) \right] 
\label{eq:ExpandedEffectiveActionKeldyshRepresentation}
\end{align}
As discussed in the paragraph following Eq.~\eqref{eq:MeanFieldEquations}, the additional degrees of freedom are a consequence of the mapping from abstract contour- to physical real-time. The second line defines a compact notation, where the Keldysh degrees of freedom are indicated by bold symbols. While $\boldsymbol{\Delta a}_{\mu}^{\alpha}$ and $\boldsymbol{\mathcal{A}}_{\mu}^{\alpha}$ are $2$-dimensional vectors in Keldysh space with ``classical'' and ``quantum'' component, $\boldsymbol{\Pi}_{\alpha \beta}^{\mu \nu}$ and $\boldsymbol{C}_{\alpha \beta}^{\mu \nu}$ are triangular $2 \times 2$ matrices with retarded, advanced and Keldysh components. The latter contain the statistical information of the theory. Since in this article we are only interested in the linear response regime at finite temperatures, the Keldysh components $(\Pi^K)_{\alpha \beta}^{\mu \nu}$ and $(C^K)_{\alpha \beta}^{\mu \nu}$ both obey the bosonic fluctuation-dissipation theorem.~\cite{KamenevBook, Note4}

In Eq.~\eqref{eq:ExpandedEffectiveActionKeldyshRepresentation} $\boldsymbol{\Pi}_{\alpha \beta}^{\mu \nu}(x, y)$ is the one-loop fermionic polarization tensor
\begin{align}
\boldsymbol{\Pi}_{\alpha \beta}^{\mu \nu}(x, y) = - \frac{i}{2} \frac{\boldsymbol{\delta}^2}{\boldsymbol{\delta a}_{\nu}^{\beta}(y) \boldsymbol{\delta a}_{\mu}^{\alpha}(x)} \textrm{tr ln} \, \boldsymbol{\hat{G}}_0^{-1} [e \boldsymbol{A}_{\mu} + \boldsymbol{a}_{\mu}^{\alpha}] \bigg|_{\boldsymbol{a} = \boldsymbol{\bar{a}}} \,,
\label{eq:DefinitionPolarizationTensor}
\end{align}
in which $\boldsymbol{\hat{G}}_0^{-1} [e \boldsymbol{A}_{\mu} + \boldsymbol{a}_{\mu}^{\alpha}]$ is the inverse fermionic propagator~\eqref{eq:InverseFreeContourTimePropagatorFlavourSpaceStructure} mapped to Keldysh space.
%
%
Referring to App.~\ref{sec:FermionicOneLoopPolarizationTensor} for details, we calculated this tensor at nonvanishing temperatures. Since the free propagators are diagonal in the flavour index, we find that the polarization tensor is diagonal in flavour space as well $\boldsymbol{\Pi}_{\alpha \beta}^{\mu \nu} = \boldsymbol{\Pi}_{(\alpha)}^{\mu \nu} \delta_{(\alpha) \beta}$. However, this may need not be the case at higher orders, so for now we keep both indices. The tensor $\boldsymbol{C}_{\alpha \beta}^{\mu \nu}(x, y)$, which we refer to as Chern-Simons--Coulomb tensor, is the integral kernel of $S_{CS}[\boldsymbol{\Delta a}_{\mu}^{\alpha}] + S_V[\boldsymbol{\bar{a}}_{\mu}^{\alpha} + \boldsymbol{\Delta a}_{\mu}^{\alpha}]$. Its real space representation reads
\begin{equation}
(C^{R/A})_{\alpha \beta}^{\mu \nu}(x, y) = (\hat{\mathcal{K}}^{-1})_{\alpha \beta} \varepsilon^{\mu \lambda \nu} \delta(x - y) \stackrel{\rightarrow}{\partial}_{\lambda} - \stackrel{\leftarrow}{\partial}_{\mu_1} \varepsilon^{0 \mu_1 \mu} (\hat{\mathcal{K}}^{-1})_{\alpha \alpha_1} V^{\alpha_1 \beta_1}(x - y) (\hat{\mathcal{K}}^{-1})_{\beta_1 \beta} \varepsilon^{0 \nu_1 \nu} \stackrel{\rightarrow}{\partial}_{\nu_1} \,,
\label{eq:RetardedAdvancedChernSimonsCoulombKernel}
\end{equation}
\end{widetext}
where the arrows above the partial derivatives indicate the direction in which they operate.

Both the fermionic polarization tensor $\boldsymbol{\Pi}_{\alpha \beta}^{\mu \nu}$, as well as the Chern-Simons--Coulomb tensor $\boldsymbol{C}_{\alpha \beta}^{\mu \nu}$ are transverse, which may be expressed by the identities
\begin{subequations}
\begin{align}
\stackrel{\rightarrow}{\partial}_{\mu} \boldsymbol{\Pi}_{\alpha \beta}^{\mu \nu} &= 0 \,, \quad \boldsymbol{\Pi}_{\alpha \beta}^{\mu \nu} \stackrel{\leftarrow}{\partial}_{\nu} = 0 \,, 
\label{eq:TransversalityConditionA} \\ \stackrel{\rightarrow}{\partial}_{\mu} \boldsymbol{C}_{\alpha \beta}^{\mu \nu} &= 0 \,, \quad \boldsymbol{C}_{\alpha \beta}^{\mu \nu} \stackrel{\leftarrow}{\partial}_{\nu} = 0 \,.
\label{eq:TransversalityConditionB}
\end{align}
\end{subequations}
As is well-known, this property is a consequence of gauge invariance.~\cite{PeskinSchroederBook} Furthermore, for the polarization tensor it is possible to factorize its tensorial structure and expand it into three distinct scalar kernels $\boldsymbol{\Pi}_{\alpha \beta}^0$, $\boldsymbol{\Pi}_{\alpha \beta}^1$ and $\boldsymbol{\Pi}_{\alpha \beta}^2$.~\cite{LopezFradkin1991, LopezFradkin1995} Transforming to Fourier space, the $2+1$-dimensional representation of this expansion, where time-like ($\mu, \nu = 0$) and space-like ($\mu, \nu = i, j = 1, 2$) indices  are separated, reads
%
\begin{subequations}
\label{eq:PolarizationTensor2+1Representation}
\begin{align}
\boldsymbol{\Pi}_{\alpha \beta}^{00}(\omega, \vec{q}) = &- \vec{q}^2 \boldsymbol{\Pi}_{\alpha \beta}^0(\omega, \vec{q}) \,,
\label{eq:PolarizationTensor2+1RepresentationA} \\
\boldsymbol{\Pi}_{\alpha \beta}^{0i}(\omega, \vec{q}) = &- \omega q^i \boldsymbol{\Pi}_{\alpha \beta}^{0}(\omega, \vec{q}) + i \varepsilon^{0ij} q_j \boldsymbol{\Pi}_{\alpha \beta}^{1}(\omega, \vec{q}) \,,
\label{eq:PolarizationTensor2+1RepresentationB} \\
\boldsymbol{\Pi}_{\alpha \beta}^{i0}(\omega, \vec{q}) = &- \omega q^i \boldsymbol{\Pi}_{\alpha \beta}^{0}(\omega, \vec{q}) - i \varepsilon^{0ij} q_j \boldsymbol{\Pi}_{\alpha \beta}^{1}(\omega, \vec{q}) \,,
\label{eq:PolarizationTensor2+1RepresentationC} \\
\boldsymbol{\Pi}_{\alpha \beta}^{ij}(\omega, \vec{q}) = &- \omega^2 \delta^{ij} \boldsymbol{\Pi}_{\alpha \beta}^{0}(\omega, \vec{q}) + i \varepsilon^{0ij} \omega \boldsymbol{\Pi}_{\alpha \beta}^{1}(\omega, \vec{q}) \nonumber \\
&+ (\delta^{ij} \vec{q}^2 - q^i q^j) \boldsymbol{\Pi}_{\alpha \beta}^{2}(\omega, \vec{q}) \,.
\label{eq:PolarizationTensor2+1RepresentationD}
\end{align}
\end{subequations}
One can readily check that the above expansion fulfills the transversality condition~\eqref{eq:TransversalityConditionA}. 

%

We close this section by a short discussion about the (anomalous) integer quantum Hall effect in graphene. Although the electromagnetic response of the interacting system to an external perturbation is encoded in the electromagnetic response tensor to be derived in the next section, the response properties of the noninteracting system are already contained in the fermionic polarization tensor $\boldsymbol{\Pi}_{\alpha \beta}^{\mu \nu}$. In fact, it is established that the essential physics of the integer quantum Hall effect can largely be understood within a noninteracting model and interactions only play a minor role.~\cite{Note5}
We therefore only need to consider $\boldsymbol{\Pi}_{\alpha \beta}^{\mu \nu}$, in particular its retarded component. The $dc$ conductivity tensor per fermionic flavour $\alpha$ can be obtained as the limit~\cite{GorbarGusynin2002}
\begin{equation}
(\sigma_0)_{\alpha}^{ij} = \lim_{\omega \to 0} \lim_{\vec{q} \to 0} \frac{e^2}{i \omega} (\Pi^R)_{\alpha}^{ij}(\omega, \vec{q}) \,,
\label{eq:ConductivityTensor}
\end{equation}
where $i,j$ are the aforementioned space-like indices ($i,j = 1,2$). Recall that to one-loop order the polarization tensor is diagonal in flavour space, hence we dropped the second flavour index.
Furthermore, we here concentrate on the off-diagonal Hall conductivity, which reduces Eq.~\eqref{eq:ConductivityTensor} to the kernel $(\Pi^R)_{\alpha}^1(0, 0)$
\begin{align}
(\sigma_0^{xy})_{\alpha} = e^2 (\Pi^R)_{\alpha}^{1}(0, 0) = \textrm{sign} \big(e B_{\textrm{eff}}^{\alpha} \big) \frac{e^2}{2 \pi} \nu_{\alpha} \,.
\label{eq:Pi1KernelLimit}
\end{align}
The second equality follows after a lengthy, but straightforward calculation. As we mentioned earlier in this section, at large magnetic fields and zero temperatures the filling fraction $\nu_{\alpha}$ is quantized into plateaus located at $\pm (n_{\alpha} + \tfrac{1}{2})$, with $n_{\alpha} = 0, 1, 2, \ldots$, see Eq.~\eqref{eq:FlavourFillingFraction} and Fig.~\ref{fig:HallPlots}. Consequently, Eq.~\eqref{eq:Pi1KernelLimit} describes the anomalous integer quantum Hall effect of the single fermionic flavour $\alpha$. A summation of the remaining flavour index then yields the Hall conductivity of the entire system of Dirac particles. For simplicity we may assume the absence of Zeeman terms and flux-binding for the moment by setting $\mu_{\alpha} = \mu$ and $B_{\textrm{eff}}^{\alpha} = B$. In that case the contributions from the individual flavours are identical, giving rise to the well-known factor of four after summing over all flavours. Restoring $\hbar$ we, thus, obtain the anomalous integer quantization of the Hall conductivity in graphene~\cite{GorbarGusynin2002, GusyninSharapov2005}
\begin{equation}
\sigma_0^{xy} = \pm \textrm{sign} \big(e B \big) \frac{e^2}{2 \pi \hbar} 4 \left( n + \frac{1}{2} \right) \,, \quad n = 0, 1, 2, \ldots \,.
\label{eq:IntegerQuantumHallConductivity}
\end{equation}
A finite temperature leads to a smearing of these plateaus, due to the thermal broadening of the Landau levels. However, since the Landau levels are not equidistant in energy because of the linear Dirac spectrum, even a small temperature eventually washes out the plateau structure at large fillings. Only the lowest levels are relatively robust against the thermal smearing. Taking into account the rather large value of the relativistic cyclotron frequency $\omega_{\textrm{c}}$, it is possible to observe the quantum Hall effect experimentally at room temperature.~\cite{Novoselov2007} By now this is a well-known fact, but still it is insofar astonishing, as the quantum Hall effect for ordinary, nonrelativistic fermions can only be observed at low temperatures, close to absolute zero.



\section{Electromagnetic Response Tensor and Hall Conductance}
\label{sec:ElectromagneticResponseTensorAndHallConductance}
In order to obtain the electromagnetic polarization tensor we need to perform the residual functional integration over the statistical gauge fields, which - according to the rules of Gaussian integration - involves the inverse of $(\boldsymbol{\Pi} + \boldsymbol{C})_{\alpha \beta}^{\mu \nu}$. However, since both $\boldsymbol{\Pi}_{\alpha \beta}^{\mu \nu}$ and $\boldsymbol{C}_{\alpha \beta}^{\mu \nu}$ are transverse, neither their individual inverse nor the inverse of their sum does exist. This problem is rooted in the gauge invariance of the partition function~\eqref{eq:DefinitionContourTimeGeneratingFunctional}. As advertized at the end of Sec.~\ref{sec:AbelianChernSimonsTheory}, we here discuss the issues of the gauge fixing procedure - resorting to the contour-time representation for the moment - and derive the electromagnetic response tensor, from which we obtain the $dc$ Hall conductivity.

The problematic gauge equivalent orbits, causing Eq.~\eqref{eq:DefinitionContourTimeGeneratingFunctional} to diverge, can be factorized from the nonequivalent physical field configurations by the well-known Fadeev-Popov gauge fixing procedure. Refering to Ref.~[\onlinecite{PeskinSchroederBook}] for details, we obtain the intermediate result
\begin{equation}
Z[\mathcal{A}_{\mu}^{\alpha}] = \mathcal{N} \int \mathcal{D} \Delta a \, \delta[G(\Delta a_{\mu}^{\alpha})] \, e^{iS_{\textrm{eff}}[\mathcal{A}_{\mu}^{\alpha}, \Delta a_{\mu}^{\alpha}]} \,.
\label{eq:GaugeFixedPartitionFunction}
\end{equation}
Here, the divergent integral over pure gauge fields as well as the so-called Fadeev-Popov determinant have been absorbed into the formally infinite normalization constant $\mathcal{N}$. Since it does not enter any correlation function this constant may safely be omitted.~\cite{Note6}
The functional delta distribution enforces the gauge constraint $G(\Delta a_{\mu}^{\alpha}) = 0$ within the functional integral, such that only physically inequivalent field configurations contribute to the amplitude. The gauge fixing function can be chosen at will, but for definiteness we consider the generalized Lorentz gauge condition,
\begin{equation}
G(\Delta a_{\mu}^{\alpha}) = \partial_{\mu} \Delta a_{\alpha}^{\mu}(x) - \omega(x) \,,
\label{eq:GeneralizedLorentzGauge}
\end{equation}
where $\omega(x)$ is an arbitrary function, in the remainder of this paper.

In its present form Eq.~\eqref{eq:GaugeFixedPartitionFunction} can in principle be employed to calculate the desired correlation functions, yet it is beneficial to make use of Feynman's trick of ``averaging over gauges''.~\cite{PeskinSchroederBook} Hereto one averages the partition function~\eqref{eq:GaugeFixedPartitionFunction} over different field configurations $\omega(x)$ with a Gaussian ``probability measure''. This procedure closely resembles a Gaussian disorder average of the partition function, albeit a disorder potential would couple in a different manner.~\cite{ChouSuYu1985, KamenevBook, SchwieteFinkelstein2014-1, SchwieteFinkelstein2014-2} The net result is the gauge fixed partition function
\begin{equation}
Z_{GF}[\mathcal{A}_{\mu}^{\alpha}] = \int \mathcal{D} \Delta a \, e^{iS_{\textrm{eff}}[\mathcal{A}_{\mu}^{\alpha}, \Delta a_{\mu}^{\alpha}] + i S_{GF}[\Delta a_{\mu}^{\alpha}]} \,,
\label{eq:GaugeFixedPartitionFunctionFinal}
\end{equation}
where the additional contribution in the exponent is the gauge fixing action
\begin{align}
S_{GF}[\Delta a_{\mu}^{\alpha}] &= \frac{1}{2 \xi} \int_{\mathcal{C}, x} \Big( \partial_{\mu} \Delta a_{\alpha}^{\mu}(x) \Big)^2 \nonumber \\
&\equiv \frac{1}{2} \int_{\mathcal{C}, xy} \Delta a_{\mu}^{\alpha}(x) \mathcal{G}_{\alpha \beta}^{\mu \nu}(x, y) \Delta a_{\nu}^{\beta}(x) \,.
\label{eq:GaugeFixingTerm}
\end{align}
Here, $\xi$ is a real-valued parameter, which may be chosen at will to simplify calculations. In the end, for any physical - that is gauge invariant - observable the dependence on $\xi$ has to drop out. After mapping this contour-time action to the physical real-time representation and performing the Keldysh rotation, the additional gauge fixing term effectively leads to the substitution $\boldsymbol{C}_{\alpha \beta}^{\mu \nu} \rightarrow (\boldsymbol{C} + \boldsymbol{\mathcal{G}})_{\alpha \beta}^{\mu \nu}$ in the effective action \eqref{eq:ExpandedEffectiveActionKeldyshRepresentation}. Since $\boldsymbol{\mathcal{G}}_{\alpha \beta}^{\mu \nu}$ is invertible so is the sum $(\boldsymbol{\Pi} + \boldsymbol{C} + \boldsymbol{\mathcal{G}})_{\alpha \beta}^{\mu \nu}$, resulting in a well-defined functional integral over the statistical gauge fields.

For nonvanishing source fields the residual Gaussian integration yields the generating functional of connected correlation functions~\cite{NegeleOrlandBook, KamenevBook}
\begin{align}
W[\boldsymbol{\mathcal{A}}_{\mu}^{\alpha}] &= -i \textrm{ln} Z_{GF}[\boldsymbol{\mathcal{A}}_{\mu}^{\alpha}] \nonumber \\
&= \int_{xy} \boldsymbol{\mathcal{A}}_{\mu}^{\alpha}(x) \boldsymbol{K}_{\alpha \beta}^{\mu \nu}(x,y) \boldsymbol{\mathcal{A}}_{\nu}^{\beta}(y) \,.
\label{eq:ConnectedFunctional}
\end{align}
In this expression $\boldsymbol{K}_{\alpha \beta}^{\mu \nu}(x,y)$ defines the electromagnetic polarization tensor.
Accordingly, it represents the linear electromagnetic response of the system to an external perturbation. We state its explicit form in terms of the fermionic polarization tensor and the Chern-Simons--Coulomb tensor of the preceeding section by employing a condensed matrix notation. For the moment the hat symbol not only indicates the flavour-space matrix structure, but also covers the discrete Minkowski indices $\mu, \nu$ and the continuous space-time variables $x, y$, if not stated otherwise,
\begin{align}
\boldsymbol{\hat{K}} &= \boldsymbol{\hat{\Pi}} - \boldsymbol{\hat{\Pi}} \left( \boldsymbol{\hat{\Pi}} + \boldsymbol{\hat{C}} + \boldsymbol{\hat{\mathcal{G}}} \right)^{-1} \boldsymbol{\hat{\Pi}} 
\label{eq:ElectromagneticResponseTensor}
\end{align}
In this expression matrix multiplication is defined naturally by implying summation over discrete and integration over continuous degrees of freedom. This tensor has the usual triangular Keldysh space structure, with retarded, advanced and Keldysh components~\cite{KamenevBook}
\begin{equation}
\boldsymbol{K}_{\alpha \beta}^{\mu \nu}(x,y) = \begin{pmatrix} 0 & (K^A)_{\alpha \beta}^{\mu \nu}(x, y) \\ (K^R)_{\alpha \beta}^{\mu \nu}(x, y) & (K^K)_{\alpha \beta}^{\mu \nu}(x, y) \end{pmatrix} \,.
\label{eq:ElectromagneticResponseTensorKeldyshStructure}
\end{equation}
Transforming to frequency-momentum space these components read
\begin{subequations}
\begin{align}
\hat{K}_{\omega, \vec{q}}^{R/A} &= \hat{\Pi}_{\omega, \vec{q}}^{R/A} - \hat{\Pi}_{\omega, \vec{q}}^{R/A} \left( \big(\hat{\Pi} + \hat{C} + \hat{\mathcal{G}} \big)_{\omega, \vec{q}}^{R/A} \right)^{-1} \hat{\Pi}_{\omega, \vec{q}}^{R/A} \,,
\label{eq:RetardedAdvancedElectromagneticResponseTensor} \\
\hat{K}_{\omega, \vec{q}}^K &= \textrm{coth}\left( \frac{\omega}{2T} \right) \left(\hat{K}_{\omega, \vec{q}}^R - \hat{K}_{\omega, \vec{q}}^A \right) \,.
\label{eq:KeldyshElectromagneticResponseTensor}
\end{align}
\end{subequations}
Here, the frequency and momentum dependence has been written as an index, flavour and Minkowski indices are still covered by the hat symbol. The second equation is just a manifestation of the bosonic fluctuation-dissipation theorem.

Although the electromagnetic response tensor as given by Eq.~\eqref{eq:ElectromagneticResponseTensor} contains the gauge fixing kernel $\boldsymbol{\hat{\mathcal{G}}}$ explicitly, any reference of it drops out in the final expression for $\boldsymbol{\hat{K}}$. In fact, the electromagnetic response tensor is a physical observable and, thus, has to be gauge-invariant. We have checked explicitly for a single flavour that other common choices, such as the Coulomb and axial gauge, indeed lead to the same result. As a consequence of gauge invariance the electromagnetic response tensor $\boldsymbol{\hat{K}}$ is transverse and, hence, admits the very same decomposition as the fermionic polarization tensor $\boldsymbol{\hat{\Pi}}$, see Eqs.~\eqref{eq:PolarizationTensor2+1RepresentationA}-\eqref{eq:PolarizationTensor2+1RepresentationD}. The only difference are the kernels $\boldsymbol{\hat{K}}_{0/1/2}$, which are now complicated functions of the kernels $\boldsymbol{\hat{\Pi}}_{0/1/2}$, the $\mathcal{K}$-matrix and the (Fourier transformation of the) Coulomb interaction matrix $\hat{V}(\vec{q})$. Recall that the latter is a $4 \times 4$ matrix in flavour-space, with all its components being equal to the same Coulomb interaction amplitude \eqref{eq:CoulombInteraction}; see also Eq.~\eqref{eq:NewInteraction}, the comments thereafter and Eq.~\eqref{eq:RetardedAdvancedChernSimonsCoulombKernel}. Suppressing frequency and momentum labels (the hat symbol only indicates flavour-space here), we obtain for the retarded kernels

\begin{widetext}
\begin{subequations}
\begin{align}
\hat{K}_0^R =&\,\, \hat{\mathcal{K}}^{-1} (\hat{\mathcal{D}}^{R})^{-1} \hat{\mathcal{K}}^{-1} \,,
\label{eq:ResponseKernelsA} \\
\hat{K}_1^R =&\,\, \hat{\mathcal{K}}^{-1} + \frac{1}{2} \hat{\mathcal{K}}^{-1} \bigg( \Big( \vec{q}^2 \hat{V} \hat{\mathcal{K}}^{-1} - \big( \hat{\Pi}_0^{R})^{-1} (\hat{\Pi}_1^{R} + \hat{\mathcal{K}}^{-1} \big) \Big) (\hat{\mathcal{D}}^{R})^{-1} + (\hat{\mathcal{D}}^{R})^{-1} \Big( \hat{\mathcal{K}}^{-1} \vec{q}^2 \hat{V} - \big( \hat{\Pi}_1^{R} + \hat{\mathcal{K}}^{-1} \big) (\hat{\Pi}_0^R)^{-1} \Big) \bigg) \hat{\mathcal{K}}^{-1} \,,
\label{eq:ResponseKernelsB} \\
\hat{K}_2^{R} =& - \frac{1}{\vec{q}^2} \hat{\mathcal{K}}^{-1} \Big( \vec{q}^2 \hat{V} \hat{\mathcal{K}}^{-1} - (\hat{\Pi}_0^{R})^{-1} \big( \hat{\Pi}_1^{R} + \hat{\mathcal{K}}^{-1} \big) \Big) (\hat{\mathcal{D}}^{R})^{-1} \Big( \hat{\mathcal{K}}^{-1} \vec{q}^2 \hat{V} - \big( \hat{\Pi}_1^{R} + \hat{\mathcal{K}}^{-1} \big) (\hat{\Pi}_0^{R})^{-1} \Big) \hat{\mathcal{K}}^{-1} \nonumber \\
&+ \frac{1}{\vec{q}^2} \hat{\mathcal{K}}^{-1} \Big( (\hat{\Pi}_0^{R})^{-1} - \vec{q}^2 \hat{V} + \omega^2 (\hat{\mathcal{D}}^{R})^{-1} \Big) \hat{\mathcal{K}}^{-1} \,,
\label{eq:ResponseKernelsC}
\end{align}
\label{eq:ResponseKernels}
\end{subequations}
%
%
with
\begin{equation}
\hat{\mathcal{D}}^{R/A} = - (\omega \pm i0)^2 \hat{\Pi}_0^{R/A} + \vec{q}^2 \big( \hat{\Pi}_2^{R/A} - \hat{\mathcal{K}}^{-1} \hat{V} \hat{\mathcal{K}}^{-1} \big) + \big( \hat{\Pi}_1^{R/A} + \hat{\mathcal{K}}^{-1} \big) (\hat{\Pi}_0^{R/A})^{-1} \big( \hat{\Pi}_1^{R/A} + \hat{\mathcal{K}}^{-1} \big) \,.
\label{eq:DMatrix}
\end{equation}
\end{widetext}
%
%
The advanced kernels are obtained by hermitean conjugation just as usual. We have to emphasize at this point, that - in contrast to the one-loop fermionic polarization tensor $\boldsymbol{\hat{\Pi}}$ - the electromagnetic polarization tensor $\boldsymbol{\hat{K}}$ is in general not diagonal in flavour space, but a symmetric matrix. This fact derives from the $\mathcal{K}$-matrix, which is also not necessarily diagonal, but symmetric.

The above equations, together with the results for the fermionic polarization tensor $\boldsymbol{\hat{\Pi}}$ given in App.~\ref{sec:FermionicOneLoopPolarizationTensor}, represent the main result of this work. Given a particular $\mathcal{K}$-matrix configuration, the electromagnetic polarization tensor $\boldsymbol{\hat{K}}$ contains the full information about the system's response to a weak, external electromagnetic perturbation. The kernel $\hat{K}_0^R$, when multiplied with $- \vec{q}^2$, equals the density response function, \textit{cf.} Eq.~\eqref{eq:PolarizationTensor2+1RepresentationA},
\begin{equation}
\boldsymbol{K}_{\alpha \beta}^{00}(\omega, \vec{q}) = - \vec{q}^2 \boldsymbol{K}_{\alpha \beta}^0(\omega, \vec{q}) \,,
\label{eq:DensityResponseFunction}
\end{equation}
and as such determines the dynamical screening properties, as well as the collective modes. The latter can be obtained by the roots of the denominator matrix $\hat{\mathcal{D}}^R$, Eq.~\eqref{eq:DMatrix}. Furthermore, in the zero temperature and long wavelength limit it is possible to calculate the absolute value square of the groundstate wavefunction and corrections thereof (as an expansion in $q/B$), which was shown in Ref.~[\onlinecite{LopezFradkin1992}]. The current response tensor is given by the spatial components, $\mu, \nu = 1, 2$, of the polarization tensor, encoding the information about the (dynamical) conductivity tensor. In the remainder of this article we focus on the $dc$ Hall conductivity. A further investigation of the above mentioned quantities will be left for future work.

In close analogy to the noninteracting case we need to investigate the zero frequency and momentum limit of the kernel~$\hat{K}_1^R$ to obtain the Hall conductivity. Using Eqs.~\eqref{eq:ResponseKernelsB},~\eqref{eq:DMatrix}, as well as $\lim_{\vec{q} \to 0} \vec{q}^2 \hat{V}(\vec{q}) = 0$, we obtain
\begin{align}
\hat{K}_1^R(0, 0) &= \lim_{\omega \to 0} \lim_{\vec{q} \to 0} \frac{1}{i \omega} (\hat{K}^R)^{12}(\omega, \vec{q}) \nonumber \\
&= \left( \hat{\mathcal{K}} + \Big( \hat{\Pi}_1^R(0, 0) \Big)^{-1} \right)^{-1} \,.
\label{eq:HallConductivityKernel}
\end{align}
Clearly, if $\hat{\mathcal{K}}$ is identically zero, the kernel $\hat{K}_1^R$ reduces to the noninteracting kernel $\hat{\Pi}_1^R$, leading back to the integer quantum Hall regime, Eq.~\eqref{eq:Pi1KernelLimit}. For the most general $\mathcal{K}$-matrix Eq.~\eqref{eq:HallConductivityKernel} reads
\begin{equation}
\hat{K}_1^R = \frac{1}{2 \pi} \begin{pmatrix} 2 k_1 + \frac{1}{\nu_1} & m_1 & n_1 & n_2 \\ m_1 & 2 k_2 + \frac{1}{\nu_2} & n_3 & n_4 \\ n_1 & n_3 & 2 k_3 + \frac{1}{\nu_3} & m_2 \\ n_2 & n_4 & m_2 & 2 k_4 + \frac{1}{\nu_4}
\end{pmatrix}^{-1} .
\label{eq:HallConductivityKernel2}
\end{equation}
Observe that the temperature dependence only enters via the kernel $\hat{\Pi}_1^R$, \textit{i.e.} via the filling fractions $\nu_{\alpha}$. A finite temperature does not modify the $\mathcal{K}$-matrix in any way, as it should be: Only the composite Dirac fermions, filling the effective Landau levels, are subject to thermal fluctuations, the flux-binding itself, as described by Eq.~\eqref{eq:SolutionMeanFieldEquations1}, is not influenced. Furthermore, note that we absorbed the sign of the effective magnetic field into the filling fractions $\nu_{\alpha}$. As discussed above, the kernel~$\hat{K}_1^R$ is a nondiagonal but symmetric matrix. In order to obtain the Hall conductivity one has to sum over all of its components
\begin{align}
\sigma_{xy} &= e^2 \sum_{\alpha, \beta} (\hat{K}_1^R)_{\alpha \beta} (0, 0) \,.
\label{eq:HallConductivity}
\end{align}
This fact becomes clear by taking into account that a physical electromagnetic fluctuation should couple identically to all flavours. Therefore, one has to neglect the flavour index of the source fields $\boldsymbol{\mathcal{A}}_{\mu}^{\alpha}(x)$ in Eq.~\eqref{eq:ConnectedFunctional}, which, in turn, leads to a summation over all matrix components rather than, \textit{e.g.}, taking a trace. Eq.~\eqref{eq:HallConductivity} is the simplest form of the Hall conductivity. Alternatively, our result could be written in terms of the (anomalous) integer quantum Hall conductivities of the noninteracting system, which may be slightly more complicated but possibly more appealing in physical terms. As advertized in the introduction we get Eq.~\eqref{eq:FractionalHallConductivityIntro}
\begin{equation*}
\sigma_{xy} = \sum_{\alpha} \sigma_{0, xy}^{\alpha} - \sum_{\alpha \beta} \sigma_{0, xy}^{\alpha} (\sigma_{0, xy} + \hat{\mathcal{K}}^{-1})_{\alpha \beta}^{-1} \sigma_{0, xy}^{\beta} \,.
\label{eq:FractionalHallConductivity}
\end{equation*}

Continuing the parallels with the noninteracting case, the Hall conductivity should be proportional to some filling factor $\nu_G$ (adopting here the notation of Ref.~[\onlinecite{Goerbig2011}]). This filling factor can easily be extracted from Eq.~\eqref{eq:HallConductivity}, using the equality $\sigma_{xy} = \frac{e^2}{2 \pi} \nu_G$. It is a complicated rational function of all the components of the $\mathcal{K}$-matrix and the filling factors of the individual composite fermions $\nu_{\alpha}$. Clearly, for such a large parameter space some of its input will be mapped to the exact same filling fraction $\nu_G$. In other words, several different FQH states produce the same filling fraction, respectively the same Hall conductivity. Hence, the measurement of a Hall plateau at a particular filling fraction alone does not identify a single FQH state. In order to distinguish from the theoretical side which state realizes a certain filling fraction in an actual experiment, one should estimate the energy associated to all of the states in question. In principle this should lead to a unique lowest energy state, which realizes that particular FQH plateau. In addition, one could investigate - theoretically and experimentally - the screening properties and/or collective modes of the respective states to gain a deeper understanding and potentially exclude a certain subset of states.

%
\onecolumngrid
\begin{center}
\begin{table}
\scalebox{1.12}{
\begin{tabular}{ | c | c | c | c | }
    \hline
   \# & $\mathcal{K}$-matrix & Total Filling Fraction $\nu_G$ & Gauge Symmetry \\ \hline
   1 & $\begin{pmatrix} 2 k & 2 k & 2 k & 2 k \\ 2 k & 2 k & 2 k & 2 k \\ 2 k & 2 k & 2 k & 2 k \\ 2 k & 2 k & 2 k & 2 k
\end{pmatrix}$ & $\frac{\nu_1 + \nu_2 + \nu_3 + \nu_4}{2k (\nu_1 + \nu_2 + \nu_3 + \nu_4) + 1}$ & $U(1)$ \\ \hline
   2a & $\begin{pmatrix} 2 k_1 & 2 k_1 & n & n \\ 2 k_1 & 2 k_1 & n & n \\ n & n & 2 k_2 & 2 k_2 \\ n & n & 2 k_2 & 2 k_2
\end{pmatrix}$ & $\frac{ \left( 2k_1 + \frac{1}{\nu_1 + \nu_2} \right) - n}{\left( 2k_1 + \frac{1}{\nu_1 + \nu_2} \right) \left( 2k_2 + \frac{1}{\nu_3 + \nu_4} \right) - n^2} + \frac{ \left( 2k_2 + \frac{1}{\nu_3 + \nu_4} \right) - n}{\left( 2k_1 + \frac{1}{\nu_1 + \nu_2} \right) \left( 2k_2 + \frac{1}{\nu_3 + \nu_4} \right) - n^2}$ & $U(1)_{\uparrow} \otimes U(1)_{\downarrow}$ \\ \hline
   2b & $\begin{pmatrix} 2 k_1 & n & 2 k_1  & n \\ n & 2 k_2 & n & 2 k_2 \\ 2 k_1 & n & 2 k_1 & n \\ n & 2 k_2 & n & 2 k_2
\end{pmatrix}$ & $\frac{ \left( 2k_1 + \frac{1}{\nu_1 + \nu_3} \right) - n}{\left( 2k_1 + \frac{1}{\nu_1 + \nu_3} \right) \left( 2k_2 + \frac{1}{\nu_2 + \nu_4} \right) - n^2} + \frac{ \left( 2k_2 + \frac{1}{\nu_2 + \nu_4} \right) - n}{\left( 2k_1 + \frac{1}{\nu_1 + \nu_3} \right) \left( 2k_2 + \frac{1}{\nu_2 + \nu_4} \right) - n^2}$ & $U(1)_{K_+} \otimes U(1)_{K_-}$ \\ \hline
   3a & $\begin{pmatrix} 2 k_1 & m_1 & 0 & 0 \\ m_1 & 2 k_2 & 0 & 0 \\ 0 & 0 & 2 k_3 & m_2 \\ 0 & 0 & m_2 & 2 k_4
\end{pmatrix}$ & $\sum_{i = 1, 2} \frac{ \left( 2k_i + \frac{1}{\nu_i} \right) - m_1}{\left( 2k_1 + \frac{1}{\nu_1} \right) \left( 2k_2 + \frac{1}{\nu_2} \right) - m_1^2} + \sum_{i = 3, 4} \frac{ \left( 2k_i + \frac{1}{\nu_i} \right) - m_2}{\left( 2k_3 + \frac{1}{\nu_3} \right) \left( 2k_4 + \frac{1}{\nu_4} \right) - m_2^2}$ & $U(1)^{\otimes 4}$ \\ \hline
   3b & $\begin{pmatrix} 2 k_1 & 0 & m_1 & 0 \\ 0 & 2 k_2 & 0 & m_2 \\ m_1 & 0 & 2 k_3 & 0 \\ 0 & m_2 & 0 & 2 k_4
\end{pmatrix}$ & $\sum_{i = 1, 3} \frac{ \left( 2k_i + \frac{1}{\nu_i} \right) - m_1}{\left( 2k_1 + \frac{1}{\nu_1} \right) \left( 2k_3 + \frac{1}{\nu_3} \right) - m_1^2} + \sum_{i = 2, 4} \frac{ \left( 2k_i + \frac{1}{\nu_i} \right) - m_2}{\left( 2k_2 + \frac{1}{\nu_2} \right) \left( 2k_4 + \frac{1}{\nu_4} \right) - m_2^2}$ & $U(1)^{\otimes 4}$ \\ \hline
    \end{tabular}
}
\caption{Filling fraction $\nu_G$ for three distinct $\mathcal{K}$-matrix configurations, leading to a Hall conductivity $\sigma_{xy} = \frac{e^2}{2 \pi \hbar} \nu_G$ ($\hbar$ restored). The examples $(2a)$ and $(2b)$, respectively $(3a)$ and $(3b)$, correspond to states with the same analytical properties but interchanged spin and valley degrees of freedom. The temperature dependence, contained within the composite-fermion filling fractions $\nu_{\alpha}$, is suppressed. For the singular $\mathcal{K}$-matrices (1) and (2a,b) there exists an equivalent abelian gauge theory with a reduced set of Chern-Simons fields. The associated gauge groups are shown in the last coloumn.
}
\label{Table}
\end{table}
\end{center}
%
\clearpage
\twocolumngrid
Considering the complexity of the matrix inverse in Eq.~\eqref{eq:HallConductivityKernel2} for the most general $\mathcal{K}$-matrix configuration, it becomes clear that a complete analysis of the full parameter space is highly involved. For its systematic study it is advisable to partially restrict the parameter space and collect the corresponding $\mathcal{K}$-matrix configurations into several distinct classes, which should have some overlap in their restricted parameter space. In this context, recall our discussion of singular $\mathcal{K}$-matrices in the preceeding section. 
Employing this strategy it is not only possible to explore the full parameter space eventually, but it also simplifies the identification of the underlying physics that is described by a particular class of $\mathcal{K}$-matrix configurations considerably. In the remainder of this paper we outline this strategy, concentrating on a few special cases. Those $\mathcal{K}$-matrix configurations we decided to investigate further, together with their resulting Hall conductivities are listed in Tab.~\ref{Table}.

\begin{figure}[t]
	\centering
		\includegraphics[width=0.475\textwidth]{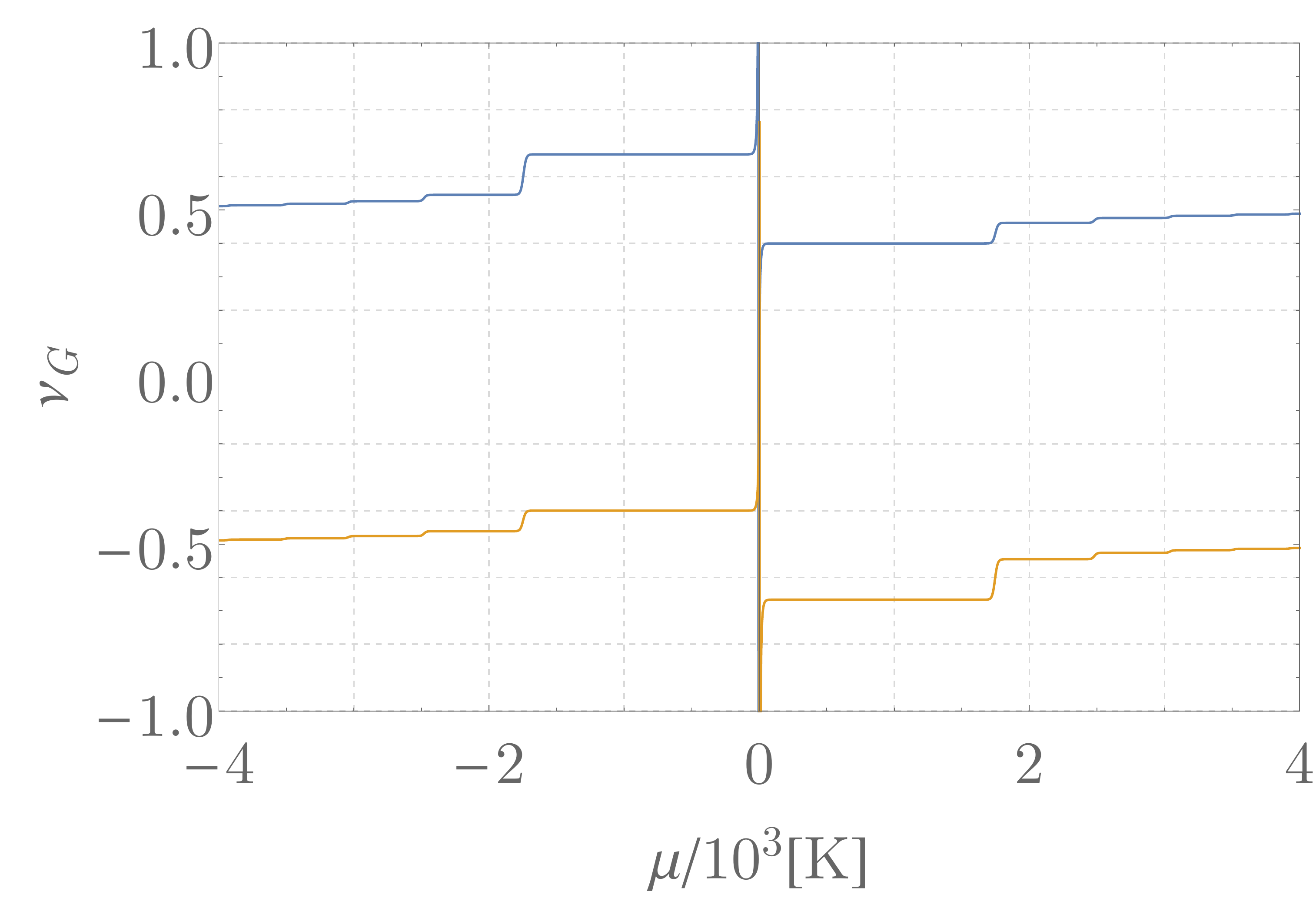}
	\caption{(Color online) Total filling fraction $\nu_G = \frac{4 \nu}{2 k \cdot 4 \nu + 1}$ as a function of the chemical potential $\mu$ at $B_{\textrm{eff}} = 15 \textrm{T}$ and $T = 10 \textrm{K}$ for $k = + 1$ (blue) and $k = - 1$ (orange). The finite temperature smears out the transitions from one Hall plateau to another, similar to the noninteracting case, \textit{cf.} Fig.~\ref{fig:HallPlots}. Note that the plateaus occur in pairs, lying symmetricaly around the charge neutrality point $\nu_G = 0$. In this simple case one can construct a manifestly particle-hole symmetric filling fraction by considering the regimes $\mu < 0$ and $\mu > 0$ and flip the sign of $k$ at $\mu = 0$, see Fig.~\ref{fig:FQHPlot2}, which yields the two branches $|\nu_G| < |1/2k|$ and $|\nu_G| > |1/2k|$.}
	\label{fig:FQHPlot1}
\end{figure}

We encountered the first of these examples already in our discussion of singular $\mathcal{K}$-matrices. The states described by this particular $\mathcal{K}$-matrix belong to the simplest possible class of FQH states, which can be described by a simpler Chern-Simons gauge theory, where only a single local $U(1)$ gauge field is present. The structure of the $\mathcal{K}$-matrix indicates a residual global $SU(4)$ flavour-symmetry, which is weakly broken by the Zeeman terms. Once the symmetry breaking terms are neglected - that is equating all composite-fermion filling fractions $\nu_{\alpha} = \nu$ - we obtain a hierarchy of states described by the filling fractions $\nu_G = \frac{4 \nu}{2 k \cdot 4 \nu + 1}$, which have also been obtained in Ref.~[\onlinecite{JellalEtal2010}]. This total filling fraction as a function of the chemical potential $\mu$ is shown in Fig.~\ref{fig:FQHPlot1} at the effective magnetic field $B_{\textrm{eff}} = 15 \textrm{T}$ and temperature $T = 10 \textrm{K}$ for $k = \pm 1$.
We remind the reader that, if one wishes to change the charge carrier density via the chemical potential, but keep the effective magnetic field $B_{\textrm{eff}}$ to be constant, then - according to the mean field equation~\eqref{eq:SelfConsistentDefinitionEffectiveMagneticField} - one has to change the external magnetic field $B$ as well. 

\begin{figure}
	\centering
		\includegraphics[width=0.475\textwidth]{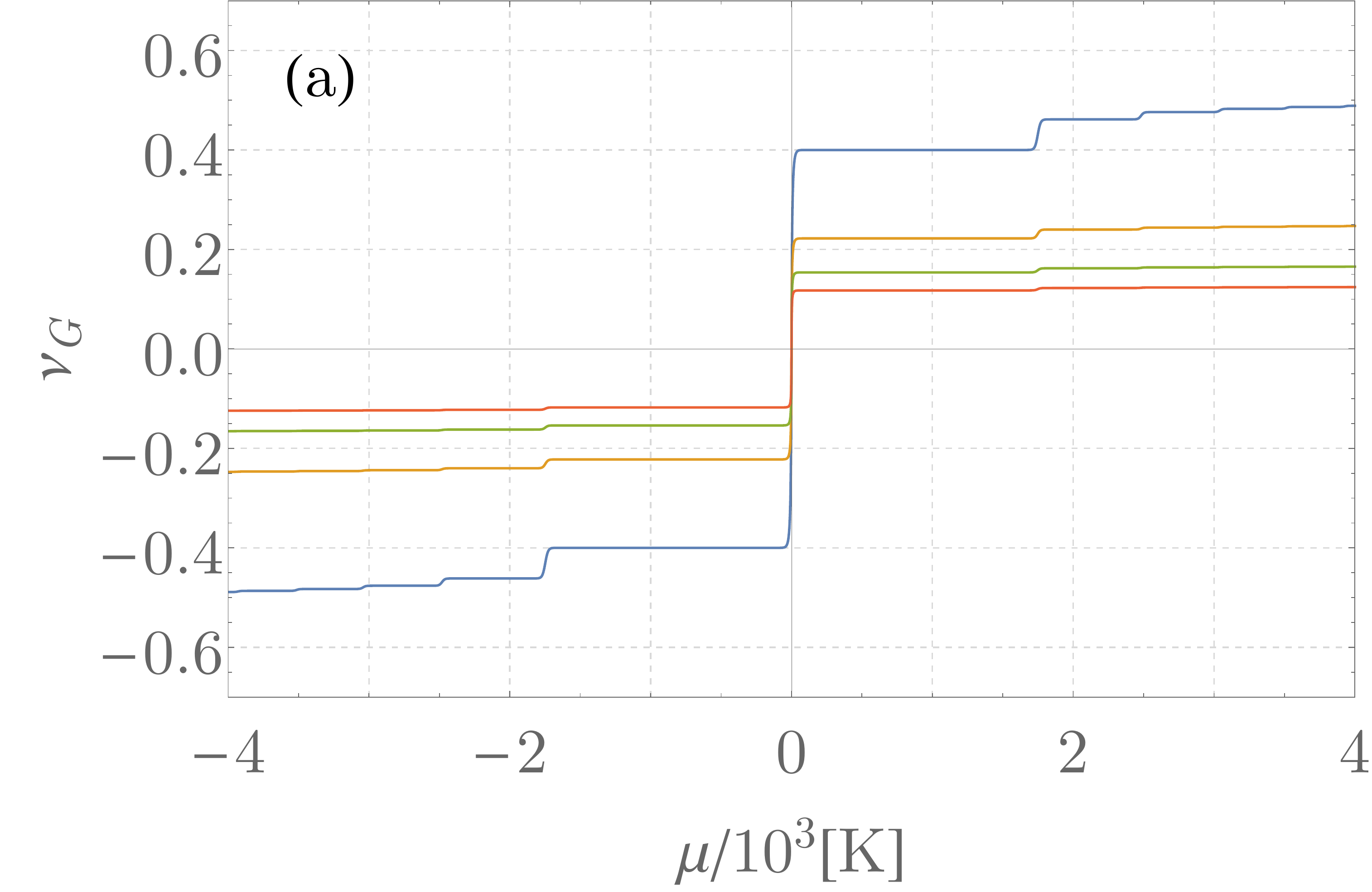} \\ \vspace{.5cm}
		\includegraphics[width=0.475\textwidth]{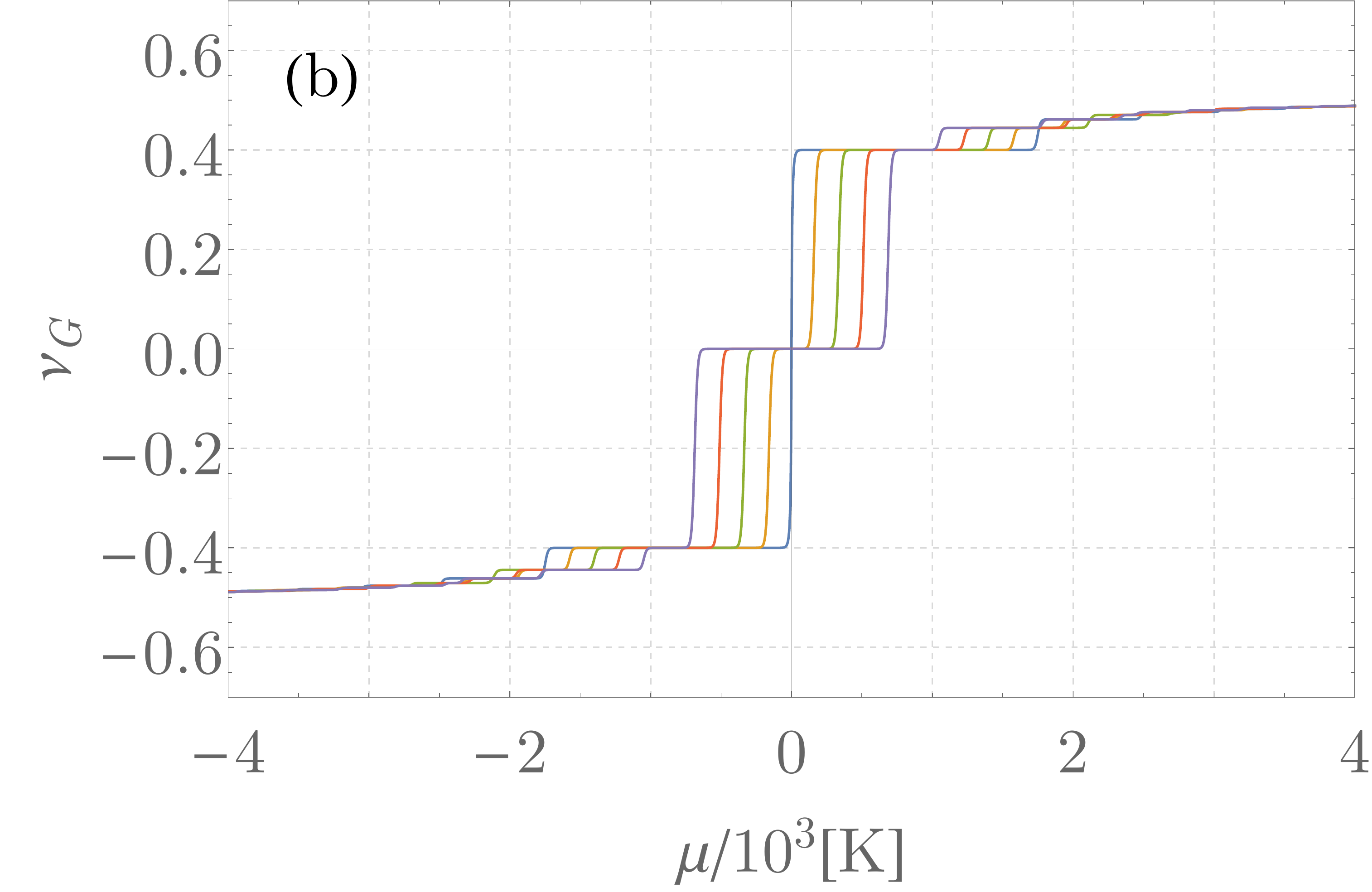}
	\caption{(Color online) Particle-hole symmetric total filling fractions for $\nu_G = \frac{(\nu_{\uparrow} + \nu_{\downarrow})}{2 k \cdot (\nu_{\uparrow} + \nu_{\downarrow}) + 1}$, with $\nu_{\uparrow} = \nu_1 + \nu_2, \nu_{\downarrow} = \nu_3 + \nu_4$, as a function of the chemical potential $\mu$ at $B_{\textrm{eff}} = 15 \textrm{T}$ and $T = 10 \textrm{K}$. (a) Zero Zeeman splitting, implying $\nu_{\uparrow} = \nu_{\downarrow} = 2 \nu$, for $|k| = 1, \ldots, 4$. (b) Finite spin Zeeman term with Zeeman energies $E_{\textrm{Z}} = 0.0, 0.1, \ldots, 0.4 \times \hbar \omega_{\textrm{c}}^{\textrm{eff}}$ for $|k| = 1$. The finite Zeeman term leads to the formation of new plateaus, with the $\nu_G = 0$ plateau being the most dominant one.}
	\label{fig:FQHPlot2}
\end{figure}

For a fixed flux-attachment prescribed by the integer $k$, it is obvious that the filling fraction $\nu_G$ is not manifestly particle-hole symmetric. Yet, the Hall plateaus occur in particle-hole symmetric pairs, when considering $k$ and $-k$ simultaneously. This observation suggests, that one can construct a manifestly particle-hole symmetric filling fraction by distinguishing the two regimes $\mu < 0$ and $\mu > 0$, and flip the sign of $k$ at $\mu = 0$, which yields the two branches
\begin{subequations}
\begin{align}
\nu_G^{\textrm{ph}} &= \frac{4 \nu}{ - 2 |k| \cdot 4 \nu + 1} \Theta(- \mu) + \frac{4 \nu}{2 |k| \cdot 4 \nu + 1} \Theta(\mu) \,, 
\label{eq:ParticleHoleSymFillingFractionA} \\
\nu_G^{\textrm{ph}*} &= \frac{4 \nu}{2 |k| \cdot 4 \nu + 1} \Theta(- \mu) + \frac{4 \nu}{- 2 |k| \cdot 4 \nu + 1} \Theta(\mu) \,,
\label{eq:ParticleHoleSymFillingFractionB}
\end{align}
\end{subequations}
where $|\nu_G^{\textrm{ph}}| < |1/2k|$ and $|\nu_G^{\textrm{ph}*}| > |1/2k|$. Note that the latter branch, $\nu_G^{\textrm{ph}*}$, appears to have the wrong overall sign. (Naively one would expect the sign of the total filling fraction $\nu_G$ to coincide with the sign of $\mu$.) But recall that we absorbed the sign of the effective magnetic field into the composite Dirac fermion filling fractions $\nu_{\alpha}$, meaning that this ``wrong sign'' should be interpreted as an effective magnetic field being antiparallel to the external one. In Fig.~\ref{fig:FQHPlot2} we show the branch~\eqref{eq:ParticleHoleSymFillingFractionA} for $|k| = 1, \ldots, 4$, as well as a generalization of Eq.~\eqref{eq:ParticleHoleSymFillingFractionA} when a finite spin Zeeman coupling is present, with Zeeman energies $E_{\textrm{Z}} = 0.1, \ldots, 0.4 \times \hbar \omega_{\textrm{c}}^{\textrm{eff}}$ for $|k| = 1$.
We have chosen such large Zeeman energy scales, which vastly exceed the ones found in a realistic graphene sample,~\cite{CastroNetoGuineaNovoselovGeim2009, Goerbig2011} for demonstrational purposes to make the additional plateau structure at $\nu_G = 0$ visible. 

The examples (2) and (3) of Tab.~\ref{Table} are best discussed comparatively. Each of these examples comes in two variations, where the flux-attachment to spin and valley degrees of freedom are interchanged. Without loss of generality we may limit our comparative discussion to the (2a) and (3b) configuration, simply referring to them as (2) and (3) if not stated otherwise.

While the (2)-configuration is another important example of a singular $\mathcal{K}$-matrix, and as such can be represented in terms of a reduced Chern-Simons theory (in this case a $U(1)_{\uparrow} \otimes U(1)_{\downarrow}$), the (3)-configuration is regular. The two $\mathcal{K}$-matrix configrations represent very different physical scenarios. The states associated to (2) are the analog of the nonrelativistic bilayer FQH states found in Ref.~[\onlinecite{LopezFradkin1993}], where an additional \textit{internal} degree of freedom - the valley polarization - in each ``spin-layer'' is present. Neglecting the Zeeman couplings in the valley subspace, that is equating $\nu_1 = \nu_2$ and $\nu_3 = \nu_4$, restores the global valley $SU(2)$ symmetry.
The states associated to the (3)-configuration on the other hand, can be interpreted as two independent, decoupled ``bilayers'', one for each valley degree of freedom. Once again, the bilayer structure is formed by the spin degree of freedom, but the valley now appears as an \textit{external} degree of freedom. (For comparison, the (3a)-configuration would yield a bilayer structure formed by the valley and the spin would appear as external degree of freedom.)

The difference of internal and external valley polarization is also reflected in the filling fraction itself, as can be seen from Tab.~\ref{Table}. For simplicity we set all composite fermion filling fractions equal, $\nu_{\alpha} = \nu$, and, furthermore, we may also set $k_1 = k_2 = k$ for the (2)-configuration, and $k_{\alpha} = k$, $m_1 = m_2 = m$ for the (3)-configuration. In those cases the $\mathcal{K}$-matrix configurations (2a) and (2b), respectively (3a) and (3b) yield the same filling fraction. If the valley appears as an internal degree of freedom we obtain
\begin{equation}
\nu_G^{\textrm{int}} = \frac{4\nu}{(2k + n) 2\nu + 1} \,,
\label{eq:FillingFracInternalValleyDeg}
\end{equation}
whereas if the valley is an external degree of freedom we get
\begin{equation}
\nu_G^{\textrm{ext}} = 2 \frac{2\nu}{(2k + m) \nu + 1} \,.
\label{eq:FillingFracExternalValleyDeg}
\end{equation}
The two filling fractions coincide by setting $n = 0$ in Eq.~\eqref{eq:FillingFracInternalValleyDeg} and $m = 2k$ in Eq.~\eqref{eq:FillingFracExternalValleyDeg}. It is this special case, which has been proposed in Ref.~[\onlinecite{PeresEtal2006}]. 

Manifestly particle-hole symmetric total filling fractions can be constructed for Eqs.~\eqref{eq:FillingFracInternalValleyDeg} and \eqref{eq:FillingFracExternalValleyDeg} in the same way as was done before, but this time at $\nu_G = 0$ one has to flip the sign of $k$ and $n$, respectively $m$, simultaneously. Similarly, the other filling fractions in Tab.~\ref{Table} also lead to particle-hole symmetric Hall plateaus. (In general, sending $\hat{\mathcal{K}} \rightarrow - \hat{\mathcal{K}}$ and $\nu_{\alpha} \rightarrow - \nu_{\alpha}$ results in $\nu_G \rightarrow - \nu_G$, \textit{cf.} Eqs.~\eqref{eq:HallConductivityKernel} and \eqref{eq:HallConductivity}.)
Hence, it is expected that for those filling fractions a similar, albeit more involved construction can be performed to present them in a manifestly particle-hole symmetric form.

As a final example we show how the prominent $\nu_G = \pm 1/3$ filling fraction, which has recently been observed in an experiment,~\cite{DuEtal2009, BolotinEtal2009, Dassarma2011} arises in our theory. To produce such a filling fraction there are several possible candidates for the $\mathcal{K}$-matrix, the simplest of which is given by the (1)-configuration of Tab.~\ref{Table} upon choosing $\sum_{\alpha} \nu_{\alpha} = \pm 1$ and $k = \pm 1$. Note that this $\mathcal{K}$-matrix configuration also gives rise to the prominent filling fractions $\nu_G = \pm 2/3$ and $\nu_G = \pm 2/5$, which are obtained by setting $\sum_{\alpha} \nu_{\alpha} = \pm 2$ and $k = \pm 1$, or $k = \pm 2$ respectively. Another possible choice for the $\mathcal{K}$-matrix is the (2)-configuration, where $n$ and $k_2$ are set to zero (one may also set $k_1 = 0$ instead). In that case the total filling fraction simplifies to $\nu_G = \frac{\nu_1 + \nu_2}{2k_1( \nu_1 + \nu_2 ) + 1} + \nu_3 + \nu_4$. Choosing the composite fermion filling fractions and the remaining flux-attachment parameter $k_1$ appropriately, that is $\nu_1 + \nu_2 = \pm 1$, $k_1 = \pm 1$ and $\nu_3 + \nu_4 = 0$, yields $\nu_G = \pm 1/3$ likewise.

Lastly, the (3)-configuration can also be employed to yield a total filling fraction of one third and it seems to us that this is the analogous configuration of the one discussed in Ref.~[\onlinecite{PapicEtal2010}], which employs the conventional wavefunction approach. In this work it was argued, that, from the four spin-valley Landau levels, two are completely filled, one is completely empty and a last one is filled to one third. Taking this statement literally one can interpret it as follows: While the completely filled (empty) levels each contribute to the filling fraction with $+\frac{1}{2} (-\frac{1}{2})$, the last level should be empty to a sixth ($-\frac{1}{2} + \frac{1}{3} = - \frac{1}{6}$).
Setting $m_1 = m_2 = 0$, as well as $k_2 = k_3 = k_4 = 0$ in the (3)-configuration of Tab.~\ref{Table}, meaning that flux is attached to one flavour only, we obtain $\nu_G = \frac{\nu_1}{2k_1 \nu_1 + 1} + \nu_2 + \nu_3 + \nu_4$. Clearly, for $\nu_1 = \nu_2 = - \frac{1}{2}$, $\nu_3 = \nu_4 = \frac{1}{2}$ and $k_1 = -2$ we reproduce the above situation
$\nu_G = - \frac{1}{6} - \frac{1}{2} + \frac{1}{2} + \frac{1}{2} = \frac{1}{3}$.
However, the actual wavefunction proposed in Ref.~[\onlinecite{PapicEtal2010}] has an $(mmm)$-like structure, meaning the Jastrow factor contains an ``off-diagonal'' vortex-attachment accounting for inter-flavour correlations between two of the four flavours, which we believe is not realized by the above simple flux-attachment. Although the precise correspondence between the $\mathcal{K}$-matrix and the electron/hole wavefunction is not yet clear, since our flux-attachment scheme refers to the charge carriers rather than electrons/holes, such off-diagonal correlations between two flavours are achieved by relaxing the constraint that, say, $m_1$ vanishes. Refering to the (3a)-configuration for definiteness, one may set $k_1 = k_2 = k$, $k_3 = k_4 = 0$ and $\nu_1 = \nu_2 = \nu$. In this special case the total filling fraction becomes $\nu_G = \frac{2 \nu}{(2k + m) \nu +1} + \nu_3 + \nu_4$. If now $k = -1$ and $m = 2k - 1 = -3$ is considered (which resembles a $\mathcal{K}$-matrix that is used in the nonrelativistic Chern-Simons theory to describe a $(333)$-state~\cite{LopezFradkin1995}), and $\nu = \nu_3 = \nu_4 = 1/2$, we obtain the desired filling fraction $\nu_G = -2/3 + 1 = 1/3$.

\section{Conclusions}
\label{sec:Conclusions}
In the present work we developed a finite temperature theory for the pseudorelativistic fractional quantum Hall effect of monolayer graphene, employing the real-time Keldysh formalism in the functional integral approach. We considered a $U(1)^{\otimes 4}$ Chern-Simons gauge theory, which is minimally coupled to the system of interacting Dirac fermions. In this theory each fermionic flavour interacts with any other flavour through Coulomb interactions, in addition to an individual $U(1)$ gauge field. The latter transforms ordinary into composite Dirac fermions. After integrating the fermionic degrees of freedom we obtained an exact effective action for the gauge fields that has been analyzed in the random phase approximation. We derived the electromagnetic response tensor from which the $dc$ Hall conductivities have been extracted.

Our research could be extended into several different directions. One obvious extension concerns a more detailed analysis of the electromagnetic response tensor for the various FQH states as presented here. The density-density response, given by $K_{\alpha \beta}^{00}$, allows for an investigation of the dynamical screening properties of the system together with the spectrum of collective modes. The current-current response, given by $K_{\alpha \beta}^{ij}$, may be studied beyond the static case, which gives information about the optical conductivity $\sigma_{ij}(\omega)$. In this context we also want to mention the straightforward generalizations of the response tensor, which result from modifications of the linear and isotropic Dirac spectrum. Here we only considered a nonvanishing (generalized) Zeeman term implicitly through the flavour dependent chemical potentials $\mu_{\alpha}$. Other modifications, such as trigonal warping, anisotropies in strained graphene, or finite mass terms (gaps), could lead to interesting effects and can be obtained by adding the respective term to the noninteracting Dirac action~\eqref{eq:ActionFermionicQuadraticPart}. Note that such alterations do not invalidate our general result given by Eqs.~\eqref{eq:ResponseKernels} and \eqref{eq:DMatrix}, if the analysis is restricted to the Gaussian approximation, but would enter via a modification of the kernels $\boldsymbol{\Pi}_{\alpha \beta}^{0/1/2}$. (The kernel expansion of $\boldsymbol{K}_{\alpha \beta}^{\mu \nu}$ is based on gauge invariance and therefore exact, but a higher order expansion in gauge fluctuations prior to integration may not only be manifested in a modification of the $\boldsymbol{\hat{\Pi}}$-kernels, but also in the form of the $\boldsymbol{\hat{K}}$-kernels, Eqs.~\eqref{eq:ResponseKernels} and \eqref{eq:DMatrix}.) Only minor modifications are involved to describe spin- or valley-polarized bilayer graphene, in the limit of weak interlayer coupling.

An important aspect in the study of the integer and fractional quantum Hall effect constitutes the role of disorder.~\cite{JainBook} As is well-known, scalar potential disorder leads to a broadening of the noninteracting Landau levels, which enter the calculation of the fermionic polarization tensor and, in turn, lead to observable consequences in the electromagnetic response spectrum, such as new kinds of collective modes (typically diffusion modes). Apart from the simple scalar potential disorder, there are other types of disorder potentials, which allow scattering processes between different flavours, causing the fermionic propagators to be nondiagonal in flavour space and may even lead to another set of collective diffusion modes.~\cite{AleinerEfetov2006, OstrovskyGornyiMirlin2006, McCannAltschuler2006, KechedzhiFalko2008} Given the large variety of possible microscopic scattering channels among the different flavours of Dirac particles and the mutual interactions between the possible collective modes, the study of disorder in graphene is a highly nontrivial task. The Keldysh formulation we employed here has proven to be an efficient computational tool for these kinds of problems, as one can perform a disorder average directly on the level of the partition function, assuming that the disorder potentials are delta correlated, which results in a fermionic pseudointeraction.~\cite{SchwieteFinkelstein2014-1, SchwieteFinkelstein2014-2, KamenevBook} In contrast to the Matsubara formulation, there is no need of the replica trick and a subsequent analytical continuation. The pseudointeraction term may then be analyzed by standard techniques, such as Hubbard-Stratonovich bosonization and/or the Wilsonian/functional renormalization group.~\cite{KamenevBook, SchwieteFinkelstein2014-1, SchwieteFinkelstein2014-2}

Another particularly interesting research direction concerns the gauge group of the Chern-Simons field itself. Here we formulated an abelian $U(1)^{\otimes 4}$ CS theory, where $SU(2)^{\otimes 2}$ and $SU(4)$ invariant states only arise as a subset of all possible FQH states obtained from the $U(1)^{\otimes 4}$ theory. The symmetry of the exact theory may only be generated as a dynamical symmetry in a more elaborate calculation, going well beyond the Gaussian fluctuations around a mean-field solution. An alternative route, where the nonabelian $SU(2)^{\otimes 2}$, respectively $SU(4)$ symmetry is manifest, would be to formulate a corresponding nonabelian gauge theory.~\cite{LopezFradkin1995} Such a theory, however, would be much more difficult to analyze, due to the additional cubic gauge field term, required by gauge invariance, and propagating Fadeev-Popov ghosts, arising from gauge fixing.~\cite{PeskinSchroederBook} Nevertheless, such a model is worth studying as it may lead to interesting insights in the fractional quantum Hall effect in graphene.

As a last remark we want to point out that our Chern-Simons theory may be of use in the conventional nonrelativistic FQHE. In this context we remind the reader of Son's proposal of a pseudorelativistic theory to explain the physics of a half-filled Landau level, Ref.~[\onlinecite{Son2015}]. Naively applying our framework for a single Dirac flavour under the assumption that charge neutrality of this relativistic model maps to half-filling of the nonrelativistic one, $\nu_{NR} = \frac{1}{2} + \frac{\nu_{CDF}}{2k \nu_{CDF} + 1}$, we made an interesting observation: Not only reproduces this formula all the particle-hole symmetric filling fractions of Jain's primary sequence around half-filling, but also those filling fractions, which are found in the Haldane-Halperin hierarchy and/or Jains secondary sequence (such as 5/13, 4/11 and 7/11 for example),~\cite{Haldane1983, Halperin1984, Jain2014} as long as $k$ is restricted to be an even integer. Of course, it could very well be the case that this feature is a mere accident, but the more appealing possibility is that there is a deeper connection between our Chern-Simons framework and Son's idea than expected. In any case it is worthwhile investigating this issue.

\acknowledgements

The author wants to thank Piet Brouwer and Zhao Liu for helpful discussions and Piet Brouwer for support in the preparation of the manuscript. This work is supported by the German Research Foundation (DFG) in the framework of the Priority Program 1459 ``Graphene''.

\appendix

\section{Fermion Propagator in External Magnetic Field}
\label{sec:FermionPropagatorInExternalMagneticField}
In this first appendix we derive the noninteracting propagator of two-dimensional Dirac particles in graphene, moving in a homogeneous magnetic field at finite temperature in Keldysh basis. This propagator has already been calculated by several authors using different methods, see for example the Refs.~[\onlinecite{Schwinger1951}], [\onlinecite{Horing2009}], and [\onlinecite{Rusin2011}], but in order to make the article self-contained we present one of those calculations, adapted to our notational conventions, here again.

The problem of inverting the operator $\hat{G}_0^{-1}$ in the quadratic form~\eqref{eq:ActionFermionicQuadraticPart} is simplified by the fact that it is diagonal in flavour space; see Eq.~\eqref{eq:InverseFreeContourTimePropagatorFlavourSpaceStructure}. Therefore the propagator itself has to be flavour diagonal
\begin{equation}
\hat{G} = \textrm{diag} \left( G_{+ \uparrow}, G_{- \uparrow}, G_{+ \downarrow}, G_{- \downarrow} \right) \,,
\label{eq:FullFermionPropagator}
\end{equation}
with $G_{\alpha} = (G_{\alpha}^{-1})^{-1}$. Thus, the problem is reduced to finding the inverse of $G_{\alpha}^{-1}$, which describes the propagation of a single flavour. Slightly abusing language, we refer to the propagator for each individual flavour $G_{\alpha}$ as ``the propagator'' in what follows. Based on the results of the mean-field approximation, Eq.~\eqref{eq:SolutionMeanFieldEquations1} and \eqref{eq:SelfConsistentDefinitionEffectiveMagneticField}, we assume that each of the flavours is subject to an individual magnetic field $B_{\textrm{eff}}^{\alpha} = B + b^{\alpha}$, and we allow each flavour to be doped individually. The propagator we obtain here occurs in the derivation of the one-loop polarization tensor~\eqref{eq:DefinitionPolarizationTensor}. The latter will be derived in detail in App.~\ref{sec:FermionicOneLoopPolarizationTensor}. In order to lighten the notation a repeated flavour space index does not imply summation. Furthermore, calculations are performed in the mixed frequency-position space.

After mapping from contour to physical time and rotating to Keldysh basis the propagator obeys the triangular Keldysh structure
\begin{equation}
\boldsymbol{G}_{\alpha}(\vec{r}, \vec{r}', \varepsilon) = \begin{pmatrix} G_{\alpha}^{K}(\vec{r}, \vec{r}', \varepsilon) & G_{\alpha}^{R}(\vec{r}, \vec{r}', \varepsilon) \\ G_{\alpha}^{A}(\vec{r}, \vec{r}', \varepsilon) & 0 \end{pmatrix} \,.
\label{eq:FlavourFermionPropagatorKeldyshStructure}
\end{equation}
As mentioned in the main text we are only interested in the linear response regime at finite temperature. Hence, the fluctuation-dissipation theorem can be employed to express the Keldysh propagator as
\begin{equation}
G_{\alpha}^{K}(\vec{r}, \vec{r}', \varepsilon) = \textrm{tanh} \left( \frac{\varepsilon}{2T} \right) \left( G_{\alpha}^{R}(\vec{r}, \vec{r}', \varepsilon) - G_{\alpha}^{A}(\vec{r}, \vec{r}', \varepsilon) \right) \,.
\label{eq:KeldyshPropagators}
\end{equation}

The retarded and advanced propagators will be constructed from the exact solutions of the stationary Dirac equation. Working in Landau gauge with the effective vector potential $\vec{A}_{\textrm{eff}}^{\alpha}(\vec{r}) = \big(- B_{\textrm{eff}}^{\alpha} y, 0 \big)^{\intercal}$, these solutions read
\begin{widetext}

\begin{subequations}
\begin{align}
\Psi_{\alpha, k_x}^0(x, \xi) &= e^{i k_x x} \begin{pmatrix} 0 \\ \psi_0(\xi) \end{pmatrix} \,, \quad \Psi_{\alpha, k_x n}^{\lambda}(x, \xi)  = \frac{1}{\sqrt{2}} e^{i k_x x} \begin{pmatrix} - \lambda \kappa_{\alpha} \psi_{n}(\xi) \\ \psi_{n + 1}(\xi) \end{pmatrix} \quad \textrm{if} \quad e B_{\textrm{eff}}^{\alpha} < 0 \,,
\label{eq:SolutionsEigenvalueEquationA} \\
\Psi_{\alpha, k_x}^0(x, \xi) &= e^{i k_x x} \begin{pmatrix} \psi_0(\xi) \\ 0 \end{pmatrix} \,, \quad \Psi_{\alpha, k_x n}^{\lambda}(x, \xi)  = \frac{1}{\sqrt{2}} e^{i k_x x} \begin{pmatrix} \psi_{n + 1}(\xi) \\ + \lambda \kappa_{\alpha} \psi_{n}(\xi) \end{pmatrix} \quad \textrm{if} \quad e B_{\textrm{eff}}^{\alpha} > 0 \,.
\label{eq:SolutionsEigenvalueEquationB}
\end{align}
\end{subequations}
where $k_x$ is a momentum quantum number, $n$ is a positive integer including zero, $\xi = \tfrac{y}{\ell_{\alpha}} + \textrm{sign}(e B_{\textrm{eff}}^{\alpha} ) k_x \ell_{\alpha}$ is a dimensionless real-space coordinate, and
$\ell_{\alpha} = \frac{1}{\sqrt{|e B_{\textrm{eff}}^{\alpha}|}}$ is the magnetic length associated to the effective magnetic field $B_{\textrm{eff}}^{\alpha}$. The spinor $\Psi_{\alpha, k_x}^0(x, \xi)$ is the zero energy Landau level located at the Dirac point, and $\Psi_{\alpha, k_x n}^{\lambda}(x, \xi)$ are Landau levels in the conduction ($\lambda = +1$) and valence band ($\lambda = -1$), respectively, whose spectrum is symmetric around the Dirac point. Recall that $\kappa_{\alpha} = \pm 1$ in the definition of the above spinors refers to the valleys $K_{\pm}$. Furthermore, $\psi_n(\xi)$ are the normalized harmonic oscillator wavefunctions
\begin{equation}
\psi_n(\xi) = \frac{1}{\sqrt{2^n n!}} \frac{1}{\pi^{1/4}} e^{- \frac{1}{2} \xi} H_n(\xi) \,,
\label{eq:HarmonicOscillatorWaveFunction}
\end{equation}
with $H_n(\xi)$ being the Hermite polynomial of degree $n$. 

In terms of the above exact solution the retarded and advanced propagators admit the following spectral decomposition
\begin{equation}
G_{\alpha}^{R/A}(\vec{r}, \vec{r}', \varepsilon) = \frac{1}{\ell_{\alpha}} \int \frac{d k_x}{2\pi} \left[ \frac{\Psi_{\alpha, k_x}^0(x, \xi) \Psi_{\alpha, k_x}^{0 \dagger}(x', \xi')}{\varepsilon + \mu_{\alpha} \pm i0} + \sum_{\lambda = \pm} \sum_{n = 0}^{\infty} \frac{\Psi_{\alpha, k_x n}^{\lambda}(x, \xi) \Psi_{\alpha, k_x n}^{\lambda \dagger}(x', \xi')}{(\varepsilon + \mu_{\alpha} \pm i0) - \lambda \sqrt{n +1} \omega_{\textrm{c}}^{\alpha}} \right] \,,
\label{eq:SpectralExpansionRetardedAdvancedFlavourPropagator}
\end{equation}
with the cyclotron frequency $\omega_{\textrm{c}}^{\alpha} = \frac{\sqrt{2} v_F}{\ell_{\alpha}}$. The momentum integration therein can be performed analytically with the help of the integral identity (Ref.~[\onlinecite{GradshteynBook}], Eq.~7.377),
\begin{equation}
\int_x e^{-x^2} H_m(y + x) H_n(z + x) = 2^n \sqrt{\pi} m! z^{n - m} L_m^{n - m}(-2yz) \,, \quad m \leq n \,.
\label{eq:IntegralIdentityHermitePolynomials}
\end{equation}
Here $L_n^k(x)$ are the associated Laguerre polynomials of degree $n$. As a result of the momentum integration, we find that the propagators can be written as a product of a translation- and gauge noninvariant phase $\chi_{\alpha}(\vec{r}, \vec{r}') = -e \int_{\vec{r}}^{\vec{r}'} \vec{A}_{\alpha}(\vec{r}'') \cdot d\vec{r}''$ - which is nothing but a Wilson line - and a translation- and gauge invariant part $S_{\alpha}^{R/A}(\vec{r} - \vec{r}', \varepsilon)$
\begin{equation}
G_{\alpha}^{R/A}(\vec{r}, \vec{r}', \varepsilon) = e^{i \chi_{\alpha}(\vec{r}, \vec{r}')} S_{\alpha}^{R/A}(\vec{r} - \vec{r}', \varepsilon) \,.
\label{eq:SolutionStructureRetardedAdvancedFlavourPropagators}
\end{equation}
Introducing the relative coordinate $\Delta \vec{r} = \vec{r} - \vec{r}'$, and the projection operators
\begin{equation}
\mathcal{P}_{\pm} = \frac{1}{2}\Big( \sigma_0 \pm \textrm{sign}\left( e B_{\alpha}^{\textrm{eff}} \right) \sigma_3 \Big) \,,
\label{eq:ChiralProjectors}
\end{equation}
the translation- and gauge invariant part of the propagators can be written compactly as
\begin{align}
S_{\alpha}^{R/A}(\Delta \vec{r}, \varepsilon) = \frac{\textrm{exp} \left(- \frac{\Delta \vec{r}^2}{4 \ell_{\alpha}^2} \right)}{4 \pi \ell_{\alpha}^2} \sum_{n = 0}^{\infty} \sum_{\lambda = \pm 1} \left[\mathcal{P}_+ L_{n}^0 \left(\frac{\Delta \vec{r}^2}{2 \ell_{\alpha}^2} \right) + \mathcal{P}_- L_{n-1}^0 \left(\frac{\Delta \vec{r}^2}{2 \ell_{\alpha}^2} \right) + i \frac{\lambda \kappa}{\sqrt{2} \ell_{\alpha}} \frac{\vec{\sigma} \cdot \Delta \vec{r}}{\sqrt{n}} L_{n-1}^1 \left(\frac{\Delta \vec{r}^2}{2 \ell_{\alpha}^2} \right) \right] S_{\alpha, \lambda n}^{R/A}(\varepsilon) \,,
\label{eq:TranslationInvariantRetardedAdvancedFlavourPropagator}
\end{align}
%
%
with
\begin{equation}
S_{\alpha, \lambda n}^{R/A}(\varepsilon) = \frac{1}{(\varepsilon + \mu_{\alpha} \pm i0) - \lambda \sqrt{n} \omega_{\textrm{c}}^{\alpha}} \,.
\label{eq:RetardedAdvancedChiralFlavourPropagators}
\end{equation}
Here we have defined $L_{-1}^0, L_{-1}^1 \equiv 0$.

The charge carrier $3$-current per flavour, $\bar{j}_{\alpha}^{\mu}$, is given by
\begin{align}
\bar{j}_{\alpha}^{\mu}(\vec{r}, t) &= - \frac{i}{2} \textrm{tr} \, \sigma_{\alpha}^{\mu} G_{\alpha}^K (\vec{r}, t, \vec{r}, t) \nonumber \\
&= - \frac{i}{2} \int_{\varepsilon} \textrm{tanh}\left( \frac{\varepsilon}{2T} \right) \textrm{tr} \, \sigma_{\alpha}^{\mu} \left( G_{\alpha}^R(\vec{r}, \vec{r}, \varepsilon) - G_{\alpha}^A(\vec{r}, \vec{r}, \varepsilon) \right) \,.
\label{eq:ChargeCarrierCurrent}
\end{align}
In thermal equilibrium only its zero component, being the charge carrier density, $\bar{j}_{\alpha}^{0} \equiv \bar{n}_{\alpha}$, acquires a finite value
\begin{equation}
\bar{n}_{\alpha}(\vec{r}, t) = \frac{1}{2 \pi \ell_{\alpha}^2} \nu_{\alpha} \,.
\label{eq:ChargeCarrierDensityAppendix}
\end{equation}
Here, $\nu_{\alpha}$ defines the filling fraction per flavour as a function of the chemical potential $\mu_{\alpha}$, the effective magnetic field $B_{\textrm{eff}}^{\alpha}$ and temperature $T$
\begin{equation}
\nu_{\alpha} = \frac{1}{2} \left[ \textrm{tanh} \left( \frac{\mu_{\alpha}}{2T} \right) + \sum_{n = 1}^{\infty} \left( \textrm{tanh} \left( \frac{\sqrt{n} \omega_{\textrm{c}}^{\alpha} + \mu_{\alpha}}{2T} \right) + \textrm{tanh} \left( \frac{- \sqrt{n} \omega_{\textrm{c}}^{\alpha} + \mu_{\alpha}}{2T} \right) \right) \right] \,.
\label{eq:FlavourFillingFractionAppendix}
\end{equation}
Near absolute zero temperature the filling fraction is quantized into plateaus of half-integers $\nu_{\alpha} = \pm \left(n_{\alpha} + \tfrac{1}{2} \right), n_{\alpha} = 0, 1, 2, \ldots$, see Fig.~\ref{fig:HallPlots}. The anomalous additional fraction occurs due to the presence of a Landau level at charge neutrality ($\mu_{\alpha} = 0$).

\section{Fermionic One-Loop Polarization Tensor}
\label{sec:FermionicOneLoopPolarizationTensor}
In this second appendix we derive the one-loop polarization tensor for Dirac fermions experiencing a homogeneous, flavour dependent effective magnetic field $B_{\textrm{eff}}^{\alpha} = B + b^{\alpha}$. See also Ref.~[\onlinecite{PyatkovskiyGusynin2010}] for a calculation of the polarization function (the $00$-component of Eq.~\eqref{eq:DefinitionPolarizationTensorAppendix}), with which our result coincides. Displaying the Keldysh structure explicitly Eq.~\eqref{eq:DefinitionPolarizationTensor} reads
\begin{align}
\boldsymbol{\Pi}_{\alpha \beta}^{\mu \nu}(x, y) = - \frac{i}{2} \frac{\boldsymbol{\delta}^2}{\boldsymbol{\delta a}_{\nu}^{\beta}(y) \boldsymbol{\delta a}_{\mu}^{\alpha}(x)} \textrm{tr ln} \begin{pmatrix} 0 & (\hat{G}_0^A)^{-1} \\ (\hat{G}_0^R)^{-1} & -(\hat{G}_0^R)^{-1}(\hat{G}_0^K)(\hat{G}_0^A)^{-1} \end{pmatrix}[e \boldsymbol{A}_{\mu} + \boldsymbol{a}_{\mu}^{\alpha}] \bigg|_{\boldsymbol{a} = \boldsymbol{\bar{a}}} \,.
\label{eq:DefinitionPolarizationTensorAppendix}
\end{align}
%
%
Recall that $\boldsymbol{\bar{a}}$ is the field expectation value of the statistical gauge field, which possesses a classical component only. Performing the functional derivatives and evaluating the result at the mean field values of the statistical gauge fields, we obtain the following retarded, advanced and Keldysh components

\begin{subequations}
\begin{align}
\big( \Pi^{R/A} \big)_{\alpha \beta}^{\mu \nu}(x - y) = \frac{i}{2} \textrm{tr} \Big( &\sigma_{\alpha}^{\mu} S_{\alpha}^{R/A}(x - y) \sigma_{\alpha}^{\nu} S_{\alpha}^K(y - x) + \sigma_{\alpha}^{\mu} S_{\alpha}^K(x - y) \sigma_{\alpha}^{\nu} S_{\alpha}^{A/R}(y - x) \Big) \delta_{\alpha \beta} \,,
\label{eq:RetardedAdvancedPolarizationTensor} \\
\big( \Pi^K \big)_{\alpha \beta}^{\mu \nu}(x - y) = \frac{i}{2} \textrm{tr} \Big( &\sigma_{\alpha}^{\mu} S_{\alpha}^R(x - y) \sigma_{\alpha}^{\nu} S_{\alpha}^{A}(y - x) + \sigma_{\alpha}^{\mu} S_{\alpha}^A(x - y) \sigma_{\alpha}^{\nu} S_{\alpha}^{R}(y - x) \nonumber \\
+ &\sigma_{\alpha}^{\mu} S_{\alpha}^K(x - y) \sigma_{\alpha}^{\nu} S_{\alpha}^K(y - x) \Big) \delta_{\alpha \beta} \,.
\label{eq:KeldyshPolarizationTensor}
\end{align}
\end{subequations}
The repeated flavour space index $\alpha$ does \textit{not} imply summation, as was the case in App.~\ref{sec:FermionPropagatorInExternalMagneticField}. Recall that the Pauli $3$-vector therein is given by $\sigma_{\alpha}^{\mu} \equiv \left( \sigma_0, \kappa_{\alpha} v_F \sigma_1, \kappa_{\alpha} v_F \sigma_2 \right)$. First, observe that the polarization tensor is diagonal in flavour space, $\boldsymbol{\Pi}_{\alpha \beta}^{\mu \nu} = \boldsymbol{\Pi}_{\alpha}^{\mu \nu} \delta_{\alpha \beta}$, which is a consequence of the free propagator being diagonal, see Eq.~\eqref{eq:FullFermionPropagator}. Second, note that the gauge- and translation-noninvariant phase $\chi_{\alpha}(\vec{r}, \vec{r}')$ drops out, such that the polarization tensor can be expressed solely in terms of the propagators $S_{\alpha}^{R/A/K}$, proving its manifest gauge- and translation-invariance. In Fourier space the above equations for the flavour diagonal components $\boldsymbol{\Pi}_{\alpha}$ become
\begin{subequations}
\begin{align}
\big( \Pi^{R/A} \big)_{\alpha}^{\mu \nu}(\omega, \vec{q}) &= \frac{i}{2} \int_{\Delta \vec{r}} \!\! e^{- i \vec{q} \cdot \Delta \vec{r}} \int_{\varepsilon} \textrm{tr} \Big( \sigma_{\alpha}^{\mu} S_{\alpha}^{R/A}(\Delta \vec{r}, \varepsilon + \omega) \sigma_{\alpha}^{\nu} S_{\alpha}^K(- \Delta \vec{r}, \varepsilon) + \sigma_{\alpha}^{\mu} S_{\alpha}^K(\Delta \vec{r}, \varepsilon) \sigma_{\alpha}^{\nu} S_{\alpha}^{A/R}(- \Delta \vec{r}, \varepsilon - \omega) \Big) \,,
\label{eq:RetardedAdvancedPolarizationTensorMomentumSpace} \\
\big( \Pi^K \big)_{\alpha}^{\mu \nu}(\omega, \vec{q}) &= \textrm{coth}\left( \frac{\omega}{2T} \right) \left( \left( \Pi^R \right)_{\alpha}^{\mu \nu}(\omega, \vec{q}) - \left( \Pi^A \right)_{\alpha}^{\mu \nu}(\omega, \vec{q}) \right) \,.
\label{eq:KeldyshPolarizationTensorMomentumSpace}
\end{align}
\end{subequations}
Eq.~\eqref{eq:KeldyshPolarizationTensorMomentumSpace} is a manifestation of the (bosonic) fluctuation-dissipation theorem. In order to arrive at this form one has to rewrite the first line of Eq.~\eqref{eq:KeldyshPolarizationTensor} according to
$\sigma^{\mu} S_{xy}^R \sigma^{\nu} S_{yx}^{A} + \sigma^{\mu} S_{xy}^A \sigma^{\nu} S_{yx}^{R} = - \sigma^{\mu} \big( S_{xy}^R - S_{xy}^{A} \big) \sigma^{\nu} \big( S_{yx}^R - S_{yx}^{A} \big)$, which holds true because of the causality properties of the retarded and advanced propagators, \textit{cf.} Ref.~[\onlinecite{KamenevBook}]. Next, one has to employ Eq.~\eqref{eq:KeldyshPropagators} and finally make use of the identity $\textrm{tanh}(x) \textrm{tanh}(y) - 1 = \textrm{coth}(x - y) \big(\textrm{tanh}(y) - \textrm{tanh}(x) \big)$.

By substituting the propagators \eqref{eq:KeldyshPropagators} and \eqref{eq:TranslationInvariantRetardedAdvancedFlavourPropagator} into Eq.~\eqref{eq:RetardedAdvancedPolarizationTensorMomentumSpace}, the polarization tensor acquires the form
\begin{align}
\big( \Pi^{R/A} \big)_{\alpha}^{\mu \nu}(\omega, \vec{q}) = \frac{1}{32 \pi^2 \ell_{\alpha}^4} \sum_{n, n'} \sum_{\lambda, \lambda'} &\frac{\mathcal{F}_{n n'}^{\lambda \lambda'}(T, \mu_{\alpha})}{(\omega \pm i0) - \lambda \sqrt{n} \omega_{\textrm{c}}^{\alpha} + \lambda' \sqrt{n'} \omega_{\textrm{c}}^{\alpha}} \nonumber \\
\times &\int_{\Delta \vec{r}} e^{- i \vec{q} \cdot \Delta \vec{r}} e^{- \frac{\Delta \vec{r}^2}{2 \ell_{\alpha}^2}} \textrm{tr} \Big( \sigma_{\alpha}^{\mu} M_n^{\alpha}(\lambda \Delta \vec{r}) \sigma_{\alpha}^{\nu} M_{n'}^{\alpha}(- \lambda' \Delta \vec{r}') \Big) \,,
\label{eq:RetardedAdvancedPolarizationTensorMomentumSpaceFinalExpression}
\end{align}
%
%
with
\begin{equation}
\mathcal{F}_{n n'}^{\lambda \lambda'}(T, \mu_{\alpha}) = \textrm{tanh}\bigg( \frac{\lambda' \sqrt{n'} \omega_{\alpha}^c - \mu_{\alpha}}{2T} \bigg) - \textrm{tanh}\left( \frac{\lambda \sqrt{n} \omega_{\alpha}^c - \mu_{\alpha}}{2T} \right) \,,
\label{eq:TemperatureFactor}
\end{equation}
and
\begin{align}
M_n^{\alpha}(\lambda \Delta \vec{r}) = \mathcal{P}_+ L_n^0 \left( \frac{\Delta \vec{r}^2}{2 \ell_{\alpha}} \right) + \mathcal{P}_- L_{n-1}^0 \left( \frac{\Delta \vec{r}^2}{2 \ell_{\alpha}} \right) + i \frac{\lambda \kappa}{\sqrt{2} \ell_{\alpha}} \frac{\vec{\sigma} \cdot \Delta \vec{r}}{\sqrt{n}} L_{n - 1}^1 \left( \frac{\Delta \vec{r}^2}{2 \ell_{\alpha}} \right) \,.
\label{eq:MatrixPartPropagators}
\end{align}
Performing the trace for each tensor component and comparing the resulting expressions with the kernel expansion~\eqref{eq:PolarizationTensor2+1Representation}, we can extract the following scalar quantities
\begin{subequations}
\begin{align}
\big( \Pi^{R/A} \big)_{\alpha}^0 (\omega, \vec{q}) =& - \frac{1}{32 \pi^2 \ell_{\alpha}^4} \frac{1}{\vec{q}^2} \sum_{n, n'} \sum_{\lambda, \lambda'} \frac{\mathcal{F}_{n n'}^{\lambda \lambda'}(T, \mu_{\alpha})}{(\omega \pm i0) - \lambda \sqrt{n} \omega_{\textrm{c}}^{\alpha} + \lambda' \sqrt{n'} \omega_{\textrm{c}}^{\alpha}} \nonumber \\
&\qquad\qquad\qquad\quad\;\;\; \times \left( I_{n-1,n'}^0(\mathcal{Q}_{\alpha}) + I_{n,n'-1}^0(\mathcal{Q}_{\alpha}) + \frac{2 \lambda \lambda'}{\sqrt{n n'}} I_{n-1,n'-1}^1(\mathcal{Q}_{\alpha}) \right) \,, \\
\big( \Pi^{R/A} \big)_{\alpha}^1 (\omega, \vec{q}) =& - \frac{\textrm{sign}(B_{\alpha}^{\textrm{eff}})}{32 \pi^2 \ell_{\alpha}^4} \frac{v_F^2}{\omega} \sum_{n, n'} \sum_{\lambda, \lambda'} \frac{\mathcal{F}_{n n'}^{\lambda \lambda'}(T, \mu_{\alpha})}{(\omega \pm i0) - \lambda \sqrt{n} \omega_{\textrm{c}}^{\alpha} + \lambda' \sqrt{n'} \omega_{\textrm{c}}^{\alpha}} \Big( I_{n-1,n'}^0(\mathcal{Q}_{\alpha}) - I_{n',n-1}^0(\mathcal{Q}_{\alpha}) \Big) \,, \\
\big( \Pi^{R/A} \big)_{\alpha}^2 (\omega, \vec{q}) =& + \frac{1}{32 \pi^2} \sum_{n, n'} \sum_{\lambda, \lambda'} \frac{\mathcal{F}_{n n'}^{\lambda \lambda'}(T, \mu_{\alpha})}{(\omega \pm i0) - \lambda \sqrt{n} \omega_{\textrm{c}}^{\alpha} + \lambda' \sqrt{n'} \omega_{\textrm{c}}^{\alpha}} \left( \frac{2 \lambda \lambda'}{\sqrt{n n'}} \frac{v_F^2}{\ell_{\alpha}^2} \partial_{\mathcal{Q}_{\alpha}}^2\tilde{I}_{n-1,n'-1}^1(\mathcal{Q}_{\alpha}) \right) \,.
\label{eq:RetardedAdvancedPolarizationKernels}
\end{align}
\end{subequations}
Here we have defined the integral expressions
%
%
\begin{subequations}
\begin{align}
I_{n,n'}^{k}(\mathcal{Q}_{\alpha}) &= \int_{\Delta \vec{r}} e^{-i \vec{q} \cdot \Delta \vec{r}} e^{- \frac{\Delta \vec{r}^2}{2 \ell_{\alpha}^2}} \left( \frac{\Delta \vec{r}^2}{2 \ell_{\alpha}^2} \right)^{k} L_n^{k}\left( \frac{\Delta \vec{r}^2}{2 \ell_{\alpha}^2} \right) L_{n'}^{k}\left( \frac{\Delta \vec{r}^2}{2 \ell_{\alpha}^2} \right) \nonumber \\
&= 2 \pi \ell_{\alpha}^2 \mathcal{Q}_{\alpha}^{n_> - n_<} e^{- \mathcal{Q}_{\alpha}} \frac{(n_< + k)!}{n_>!} L_{n_<}^{n_> - n_<}(\mathcal{Q}_{\alpha}) L_{n_< + k}^{n_> - n_<}(\mathcal{Q}_{\alpha}) \,, \quad k = 0,1 \,,
\label{eq:DefinitionIntegralExpressionPolarizationKernelsA} \\
\tilde{I}_{n,n'}^1(\mathcal{Q}_{\alpha}) &= \int_{\Delta \vec{r}} e^{-i \vec{q} \cdot \Delta \vec{r}} e^{- \frac{\Delta \vec{r}^2}{2 \ell_{\alpha}^2}} L_n^{1}\left( \frac{\Delta \vec{r}^2}{2 \ell_{\alpha}^2} \right) L_{n'}^{1}\left( \frac{\Delta \vec{r}^2}{2 \ell_{\alpha}^2} \right) \nonumber \\
&= \sum_{m=0}^n \sum_{m'=0}^{n'} I_{m, m'}^0(\mathcal{Q}_{\alpha}) \,,
\label{eq:DefinitionIntegralExpressionPolarizationKernelsB}
\end{align}
\end{subequations}
where $\mathcal{Q}_{\alpha} = \frac{\vec{q}^2 \ell_{\alpha}^2}{2}$ is a dimensionless momentum variable, and $n_> = \textrm{max}\{n,n'\}$, $n_< = \textrm{min}\{n,n'\}$. Note that both $I_{n,n'}^k$ and $\tilde{I}_{n,n'}^1$ are symmetric in their Landau indices $n, n'$. Hence, without loss of generality we can assume $n \leq n'$ in the following proof.

First let us show how $\tilde{I}_{n,n'}^1(\mathcal{Q}_{\alpha})$ can be reduced to a sum of $I_{n,n'}^{0}(\mathcal{Q}_{\alpha})$. In order to prove this equality we only have to make use of the property $L_n^{k + 1}(x) = \sum_{m = 0}^n L_m^{k}(x)$, see Ref.~[\onlinecite{GradshteynBook}] (Eq.~8.974.3), and interchange integration and summation. We immediately arrive at the second line of Eq.~\eqref{eq:DefinitionIntegralExpressionPolarizationKernelsB}. The proof of Eq.~\eqref{eq:DefinitionIntegralExpressionPolarizationKernelsA} is more involved. First of all, one has to work in polar coordinates, substituting $t = \frac{\Delta \vec{r}^2}{2 \ell_{\alpha}^2}$, and perform the angle integration, which yields the Bessel function of the first kind $J_0$
\begin{equation}
I_{n,n'}^{k}(\mathcal{Q}_{\alpha}) = 2 \pi \ell_{\alpha}^2 \int_0^{\infty} dt \, e^{- t} t^k J_0 \left(2 \sqrt{\mathcal{Q}_{\alpha} t} \right) L_{n_>}^k(t) L_{n_<}^k(t) \,.
\label{eq:DerivationIntegralExpressions1}
\end{equation}
Next, we rewrite $L_{n_<}^k(t) = (- t)^{- k} \frac{(n_< + k)!}{n_<!} L_{n_< + k}^{- k}(t)$, see Ref.~[\onlinecite{PyatkovskiyGusynin2010}], resulting in
\begin{equation}
I_{n,n'}^{k}(\mathcal{Q}_{\alpha}) = 2 \pi \ell_{\alpha}^2 (- 1)^k (n_< + 1)^k \int_0^{\infty} dt \, e^{- t} J_0 \left(2 \sqrt{\mathcal{Q}_{\alpha} t} \right) L_{n_>}^k(t) L_{n_< + k}^{- k}(t) \,.
\label{eq:DerivationIntegralExpressions2}
\end{equation}
The residual integration can be performed by making use of the integral identity (Ref.~[\onlinecite{GradshteynBook}], Eq.~7.422.2)
\begin{equation}
\int_0^{\infty} dt \, e^{- t} J_0 \left(2 \sqrt{\mathcal{Q}_{\alpha} t} \right) L_{n_>}^k(t) L_{n_< + k}^{- k}(t) = (- 1)^{n_> + n_< + k} e^{- \mathcal{Q}_{\alpha}} L_{n_>}^{n_< - n_>}(\mathcal{Q}_{\alpha}) L_{n_< + k}^{n_> - n_<}(\mathcal{Q}_{\alpha}) \,.
\label{eq:IntegralIdentityLaguerrePolynomials}
\end{equation}
After straightforward manipulation of the result we find Eq.~\eqref{eq:DefinitionIntegralExpressionPolarizationKernelsA} eventually.

\end{widetext}



%

\end{document}